%% file: ceed.tex
\documentclass[Afour,sageh,times]{sagej}
\input tex/header

\begin{document}

\title{Efficient Exascale Discretizations: High-Order Finite Element Methods}

\author{
Tzanio Kolev\affilnum{1},
Paul Fischer\affilnum{2,3,4},
Misun Min\affilnum{2},
Jack Dongarra\affilnum{5},
Jed Brown\affilnum{6},
Veselin Dobrev\affilnum{1},
Tim Warburton\affilnum{7},
Stanimire Tomov\affilnum{5},
Mark S. Shephard\affilnum{8},
Ahmad Abdelfattah\affilnum{5},
Valeria Barra\affilnum{6},
Natalie Beams\affilnum{5},
Jean-Sylvain Camier\affilnum{1},
Noel Chalmers\affilnum{9},
Yohann Dudouit\affilnum{1},
Ali Karakus\affilnum{10},
Ian Karlin\affilnum{1},
Stefan Kerkemeier\affilnum{2},
Yu-Hsiang Lan\affilnum{2},
David Medina\affilnum{11},
Elia Merzari\affilnum{2,12},
Aleksandr Obabko\affilnum{2},
Will Pazner\affilnum{1},
Thilina Rathnayake\affilnum{3},
Cameron W. Smith\affilnum{5},
Lukas Spies\affilnum{3},
Kasia Swirydowicz\affilnum{13},
Jeremy Thompson\affilnum{6},
Ananias Tomboulides\affilnum{2,14},
Vladimir Tomov\affilnum{1}
}

\affiliation{%
\affilnum{1}Center for Applied Scientific Computing, Lawrence Livermore National Laboratory, Livermore, CA 94550\\
\affilnum{2}Mathematics and Computer Science, Argonne National Laboratory, Lemont, IL 60439\\
\affilnum{3}Department of Computer Science, University of Illinois at Urbana-Champaign, Urbana, IL 61801\\
\affilnum{4}Department of Mechanical Science and Engineering, University of Illinois at Urbana-Champaign, Urbana, IL 61801\\
\affilnum{5}Innovative Computing Laboratory, University of Tennessee, Knoxville, TN 37996\\
\affilnum{6}Department of Computer Science, University of Colorado, Boulder, CO 80309\\
\affilnum{8}Scientific Computation Research Center, Rensselaer Polytechnic Institute, Troy, NY 12180\\
\affilnum{7}Department of Mathematics, Virginia Tech, Blacksburg, VA 24061\\
\affilnum{9}AMD Research, Austin, TX 78735\\
\affilnum{10}Mechanical Engineering Department, Middle East Technical University, 06800, Ankara, Turkey\\
\affilnum{11}Occalytics LLC, Weehawken, NJ 07086\\
\affilnum{12}Department of Nuclear Engineering, Penn State, PA 16802\\
\affilnum{13}Pacific Northwest National Laboratory, WA 99352\\
\affilnum{14}Department of Mechanical Engineering, Aristotle University of Thessaloniki, Greece 54124\\
}

\corrauth{Tzanio Kolev,
Center for Applied Scientific Computing, Lawrence Livermore National Laboratory, Livermore, CA 94550}
\email{tzanio@llnl.gov}

\begin{abstract}
Efficient exploitation of exascale architectures requires rethinking of the
numerical algorithms used in many large-scale applications. These architectures
favor algorithms that expose ultra fine-grain parallelism and maximize the ratio
of floating point operations to energy intensive data movement. One of the few
viable approaches to achieve high efficiency in the area of PDE discretizations
on unstructured grids is to use matrix-free / partially-assembled high-order
finite element methods, since these methods can increase the accuracy and/or
lower the computational time due to reduced data motion. In this paper we
provide an overview of the research and development activities in the Center for
Efficient Exascale Discretizations (CEED), a co-design center in the Exascale
Computing Project that is focused on the development of next-generation
discretization software and algorithms to enable a wide range of finite element
applications to run efficiently on future hardware. CEED is a research
partnership involving more than 30 computational scientists from two US national
labs and five universities, including members of the Nek5000, MFEM, MAGMA and
PETSc projects. We discuss the CEED co-design activities based on targeted
benchmarks, miniapps and discretization libraries and our work on performance
optimizations for large-scale GPU architectures. We also provide a broad
overview of research and development activities in areas such as unstructured
adaptive mesh refinement algorithms, matrix-free linear solvers, high-order data
visualization, and list examples of collaborations with several ECP and external
applications.
\end{abstract}


\maketitle

\input tex/introduction
\input tex/codesign
\input tex/ecosystem
\input tex/benchmarks
\input tex/miniapps
\input tex/libceed
\input tex/nek+mfem
\input tex/applications
\input tex/conclusion
\input tex/acknowledgments
\input tex/funding

\bibliographystyle{SageH}
\bibliography{ceed}

\end{document}

%% file: tex/header.tex
\usepackage{moreverb,url}
\usepackage{amssymb,amsmath,amsfonts}
\usepackage{subfig}
\usepackage{graphicx}
\usepackage{rotating}
\usepackage{geometry}
\usepackage{setspace}
\usepackage{lipsum}
\newcommand{\verbatimfont}[1]{\renewcommand{\verbatim@font}{\ttfamily#1}}
\usepackage{epstopdf}
\usepackage{alltt}
\usepackage{color}
\usepackage{multirow}
\usepackage{listings}

\usepackage{fancyvrb}
\fvset{fontsize=\scriptsize}
\RecustomVerbatimEnvironment{verbatim}{Verbatim}{}

\usepackage{floatrow}
\usepackage{algpseudocode,algorithm,algorithmicx}
\usepackage{algcompatible}
\usepackage{todonotes}
\usepackage{float}

\usepackage{tikz}
\usepackage{pgfplots}

\definecolor{code}{rgb}{0, 0, 0}
\newcommand{\code}[1]{\texttt{\small\color{code} #1}}

\definecolor{RYB1}{RGB}{166,206,227}  
\definecolor{RYB2}{RGB}{31,120,180}   
\definecolor{RYB3}{RGB}{178,223,138}  
\definecolor{RYB4}{RGB}{51,160,44}    
\definecolor{RYB5}{RGB}{251,154,153}  
\definecolor{RYB6}{RGB}{227,26,28}    
\definecolor{RYB7}{RGB}{253,191,111}  
\definecolor{RYB8}{RGB}{255,127,0}    
\definecolor{RYB9}{RGB}{202,178,214}  
\definecolor{RYB10}{RGB}{160,60,140}  
\definecolor{RYB11}{RGB}{255,255,153} 
\definecolor{RYB12}{RGB}{177,89,40}   

\pgfplotscreateplotcyclelist{will}{%
black,every mark/.append style={fill=white},mark=*\\%
RYB4!50!black,every mark/.append style={fill=RYB4},mark=*\\%
RYB6!50!black,every mark/.append style={fill=RYB6},mark=*\\%
RYB8!50!black,every mark/.append style={fill=RYB8},mark=*\\%
RYB10!50!black,every mark/.append style={fill=RYB10},mark=*\\%
RYB12!50!black,every mark/.append style={fill=RYB12},mark=*\\%
black,every mark/.append style={fill=white!50!black},mark=*\\%
RYB2!50!black,every mark/.append style={fill=RYB2},mark=*\\%
RYB4!50!black,densely dashed,every mark/.append style={fill=RYB4},mark=*\\%
RYB6!50!black,densely dashed,every mark/.append style={fill=RYB6},mark=*\\%
RYB8!50!black,densely dashed,every mark/.append style={fill=RYB8},mark=*\\%
RYB10!50!black,densely dashed,every mark/.append style={fill=RYB10},mark=*\\%
RYB12!50!black,densely dashed,every mark/.append style={fill=RYB12},mark=*\\%
black,densely dashed,every mark/.append style={fill=white!50!black},mark=*\\%
RYB2!50!black,densely dashed,every mark/.append style={fill=RYB2},mark=*\\%
}

\input tex/pf

%% file: tex/pf.tex
\def\dO{\partial \Omega}

\def\scriptO{{{\it O}\kern -.42em {\it `}\kern + .20em}}
\def\RR{{{\rm l}\kern - .15em {\rm R} }}
\def\PP{{{\rm l}\kern - .15em {\rm P} }}
\def\L2{{{\sf L}^2}}
\def\H1{{{\sf H}^1}}

\def\PN2{{\PP_{N}-\PP_{N-2}}}

\def\complex{{{\rm C} \kern - .53em {\rm l} \kern + .38em}}

\def\a1{{ | \lambda_{\min} |}}

\def\l1{{   \lambda_{\min}  }}

\def\bu0{{\underline {\bf 0}}}

\def\br{{\bf r}}
\def\bu{{\bf u}}

\def\bx{{\bf x}}

\def\Oh{{\hat \Omega}}

\def\uf{{\underline f}}

\def\uu{{\underline u}}

\def\u0{{\underline 0}}
\def\n12{{n_{\frac{1}{2}}}}
\def\t12{{t_{\frac 1 2}}}
\def\n12{{n_{0.8}}}
\def\t12{{t_{0.8}}}

%% file: tex/introduction.tex
\section{Introduction} \label{sec:introduction}

Efficient exploitation of exascale architectures requires rethinking of the
numerical algorithms for solving partial differential equations (PDEs) on
general unstructured grids. New architectures, such as general purpose graphics
processing units (GPUs) favor algorithms that expose ultra fine-grain
parallelism and maximize the ratio of floating point operations to energy
intensive data movement.

Many large-scale PDE-based applications employ unstructured finite element
discretization methods, where practical efficiency is measured by the accuracy
achieved per unit computational time. One of the few viable approaches to
achieve high performance in this case is to use matrix-free high-order finite
element methods, since these methods can both increase the accuracy and/or lower
the computational time due to reduced data motion. To achieve this efficiency,
high-order methods use mesh elements that are mapped from canonical reference
elements (hexahedra, wedges, pyramids, tetrahedra) and exploit, where possible, the
tensor-product structure of the canonical mesh elements and finite element
spaces. Through matrix-free partial assembly, the use of canonical reference
elements enables substantial cache efficiency and minimizes extraneous data
movement in comparison to traditional low-order approaches.

The Center for Efficient Exascale Discretizations (CEED) is a focused team
effort within the U.S. Department of Energy (DOE) Exascale Computing Project
(ECP) that aims to develop the next-generation discretization software and
algorithms to enable a wide range of finite element applications to run
efficiently on future hardware. CEED is a research partnership involving more
than 30 computational scientists from two DOE labs and five universities,
including members of the Nek5000, MFEM, MAGMA and PETSc projects \citep{nek,
mfem, mfem-web, magma, occa, petsc-web-page}. This article provides an
overview of the co-design research and development activities in the CEED
project based on targeted benchmarks, miniapps and discretization libraries.
We also discuss several examples of collaborations with ECP, including
ExaSMR, MARBL, Urban, and ExaWind, as well as external applications.

Following the ECP co-design philosophy, CEED is positioned as a computational
motif hub between applications, hardware vendors and software technologies
projects. As such, the main objectives of the project are to:
\begin{enumerate}
\item Help applications leverage future architectures by providing them with
  state-of-the-art discretization algorithms that better exploit the hardware
  and deliver a significant performance gain over conventional low-order
  methods.
\item Collaborate with hardware vendors and software technologies projects to
  utilize and impact the upcoming exascale hardware and its software stack
  through CEED-developed proxies and miniapps.
\item Provide an efficient and user-friendly unstructured PDE discretization
  component for the upcoming exascale software ecosystem.
\end{enumerate}

To address these objectives, the center's co-design efforts are organized in
four interconnected research and development thrusts, focused on the following
computational motifs and their performance on exascale hardware:

\medskip

\noindent
{\bf PDE-based simulations on unstructured grids.} CEED is producing a range of
software products supporting general finite element algorithms on triangular,
quadrilateral, tetrahedral and hexahedral meshes. We target the whole de Rham
complex: $H^1$, $H(\mathrm{curl})$, $H(\mathrm{div})$ and $L^2$/DG spaces and
discretizations, including conforming and non-conforming unstructured adaptive
mesh refinement (AMR).

\medskip

\noindent
{\bf High-order/spectral finite elements.} Our algorithms and software come with
comprehensive high-order support: we provide efficient matrix-free operator
evaluation for any order space on any order mesh, including high-order curved
meshes and all geometries in the de Rham complex. The CEED software also
includes optimized assembly support for low-order methods.

The rest of the paper is organized as follows. In Section \ref{sec:codesign} we
describe our co-design goals and organization. The needs of a complete
high-order software ecosystem are then reviewed in Section \ref{sec:ecosystem}.
The CEED high-order benchmarks designed to test and compare the performance of
high-order codes are described in Section \ref{sec:benchmarks}. CEED is
developing a variety of miniapps encapsulating key physics and numerical kernels
of high-order applications. These are described in Section
\ref{sec:miniapps}. We deliver performant algorithms to applications via
discretization libraries both at low-level, see libCEED described in Section
\ref{sec:libceed}, and high-level, see MFEM and Nek described in Section
\ref{sec:NekMFEM}. The impact of these CEED-developed technologies in several
applications is illustrated in Section \ref{sec:applications}, followed by
conclusions in Section \ref{sec:conclusions}.

%% file: tex/codesign.tex
\section{Co-Design} \label{sec:codesign}

CEED's co-design activities are organized in four R\&D thrusts described below.

\begin{figure*}[ht]
\centering
\includegraphics[width=0.8\textwidth]{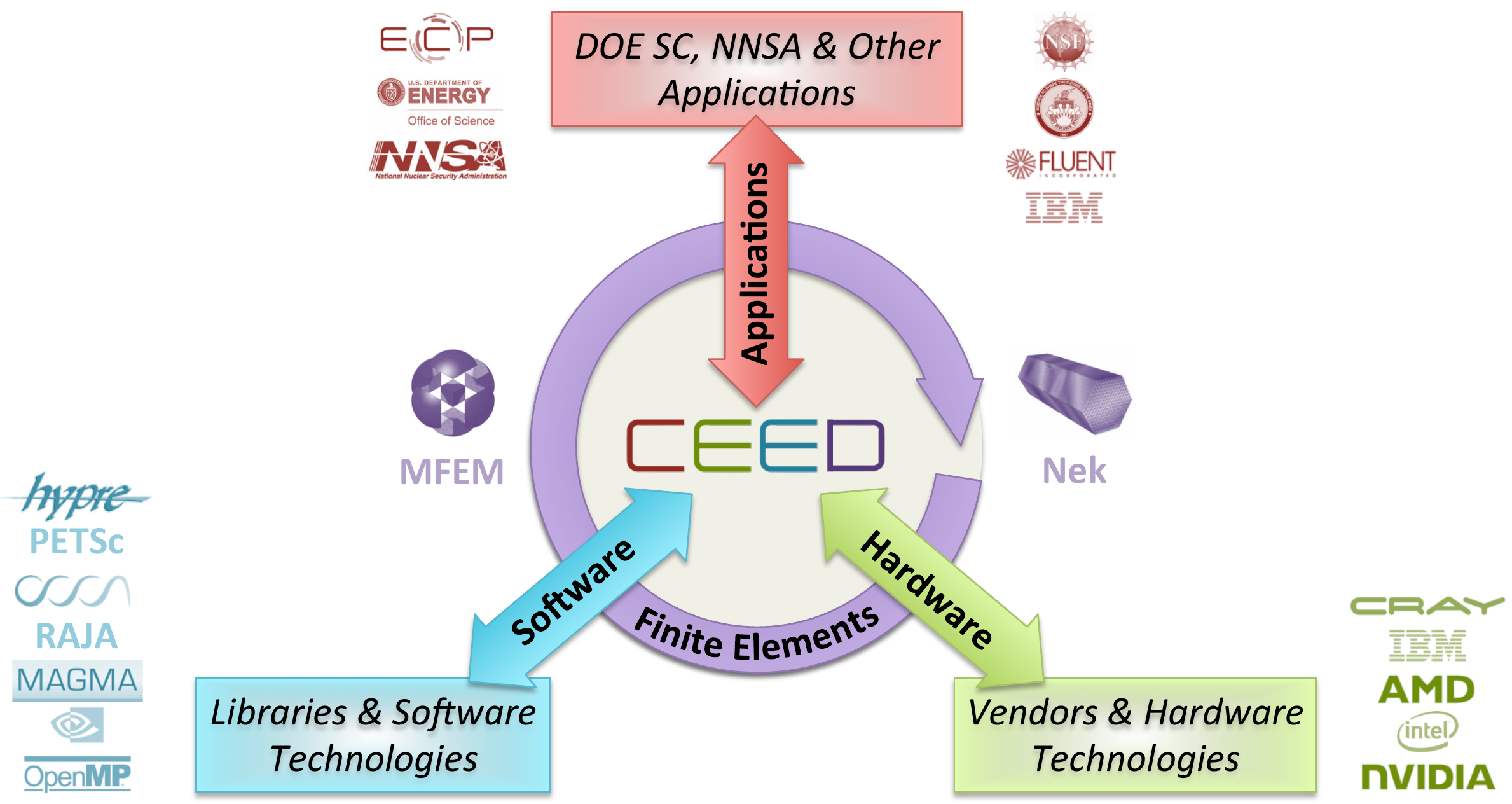}
\caption{CEED research and development thrusts}
\label{ceed-thrusts}
\end{figure*}

\noindent
{\bf Applications Thrust.} The goal of CEED's Applications thrust is to impact a
wide range of ECP application teams through focused one-on-one interactions,
facilitated by CEED application liaisons, as well as through one-to-many
interactions, based on the development of easy-to-use discretization libraries
for high-order finite element methods.

\smallskip

\noindent
{\bf Hardware Thrust.} The goal of CEED's Hardware thrust is to build a two-way
(pull-and-push) collaboration with vendors, where the CEED team will develop
hardware-aware technologies (pull) to understand performance bottlenecks and
take advantage of inevitable hardware trends, and vendor interactions to seek
(push) impact and improve hardware designs within the ECP scope.

\smallskip

\noindent
{\bf Software Thrust.} The goal of CEED's Software thrust is to participate in
the development of software libraries and frameworks of general interest to the
scientific computing community, facilitate collaboration between CEED software
packages, enable integration into and/or interoperability with overall ECP
software technologies stack, streamline developer and user workflows, maintain
testing and benchmarking infrastructure, and coordinate CEED software releases.

\smallskip

\noindent
{\bf Finite Elements Thrust.} The goal of CEED's Finite Element thrust is to
continue to improve the state-of-the-art high-order finite
element and spectral element algorithms and kernels in the CEED software targeting exascale
architectures, connect and contribute to the efforts of the other thrusts, and
lead the development of discretization libraries, benchmarks and miniapps.

\smallskip

The CEED co-design approach is driven by applications, and is based on close
collaboration between the Applications, Hardware, and Software thrusts, each of
which has a two-way, push-and-pull relation with the external application,
hardware and software technologies teams. CEED's Finite Elements thrust serves
as a central hub that ties together, coordinates and contributes to the efforts
in all thrusts. For example, the development of discretization libraries in CEED
is led by the Finite Elements thrust but involves working closely with vendors
(Hardware thrust) and software technology efforts (Software thrust) to take full
advantage of exascale hardware. Making sure that these libraries meet the needs
of, and are successfully incorporated in, ECP applications is based on
collaboration between the Applications and Finite Elements thrusts.

To facilitate the co-design process, the CEED project is developing a number of
benchmarks, libraries of highly performant kernels, and a set of miniapps that
are serving multiple roles. One of these roles is to provide a mechanism to test
and optimize across the breadth of implementations already developed by team
members for a variety of platforms. The CEED bake-off problems (BPs) described
in Section \ref{sec:benchmarks} were specifically designed for that
purpose. They are simple enough to be able to be run in a simulator, but include
the key local and global kernels in model problem settings. CEED also provides
well-documented miniapps that are simple yet capture application-relevant
physics to work with vendors, be used in system procurement, collaborate
software technologies projects, and provide test and demonstration cases for
application scientists. These miniapps, which are one step above the benchmarks
are described in Section \ref{sec:miniapps}. One of their uses is to highlight
performance critical paths (e.g. size of on package memory, internode latency,
hardware collectives) with the goal to impact the design of exascale
architectures, and system and application software, for improved portability and
performance of the high-order algorithms. All of the optimizations and
performance improvements resulting from the benchmarks and miniapps work is made
available to applications via the CEED discretization libraries described in
Sections \ref{sec:libceed} and \ref{sec:NekMFEM}.

%% file: tex/ecosystem.tex
\section{High-Order Software Ecosystem} \label{sec:ecosystem}

While the main focus of the CEED effort is the development and improvement of
efficient discretization algorithms, a full-fledged high-order application
software ecosystem requires many other components: from meshing, to adaptivity,
solvers, visualization and more. Therefore, CEED is also engaged in improving
the additional components of the overall high-order simulation pipeline. We
describe some of these efforts as well as some key enabling technologies in this
section to provide a background for the discretization work discussed in the
remainder of the paper. Note that some of the components described below (e.g.
the MAGMA and OCCA projects) are generally applicable and could be useful in
applications that do not use finite elements methods.

\subsection*{High-Order Meshing}

When applying high-order discretization methods over domains with curved
boundaries and/or curved material interfaces, the mesh must maintain a curved
mesh geometric approximation, whose order is dictated by the order of the basis
functions used to discretize the PDEs to ensure convergence of the solution. In
the case when Lagrangian reference frame methods are applied the mesh geometry
will naturally become curved to the same order as the elements discretizing the
PDEs. Thus, the application of high-order methods requires the ability to
generate curved initial meshes and to support curved mesh adaptation whenever
adaptive mesh control is applied. To meet these needs the CEED software supports
curved mesh representations and has developed tools for curved mesh adaptation
that include non-conforming mesh refinement/de-refinement of quadrilateral and
hexahedral meshes, and conforming mesh adaptation of triangular and tetrahedral
meshes that can refine and coarsen the mesh to match a given anisotropic mesh
metric field.

\begin{figure}[ht]
\centering
\includegraphics[width=0.95\linewidth]{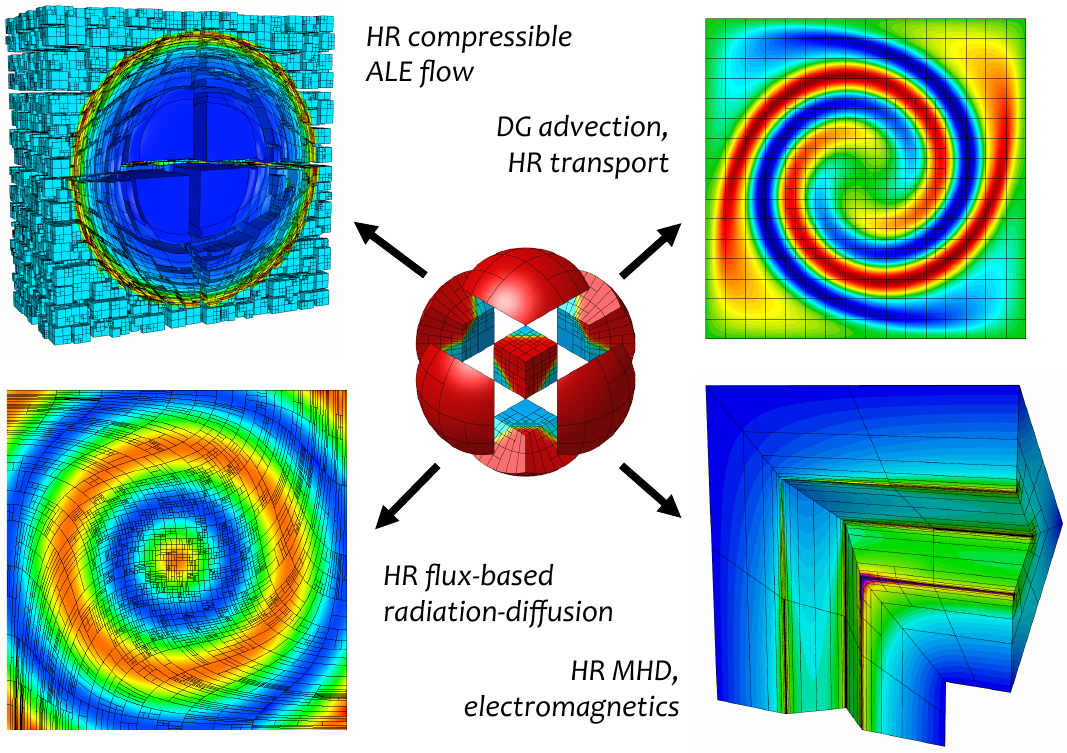}
\caption{By incorporating AMR at the library level, many MFEM-based applications
  can take advantage of it with minimal code changes. Examples from high-order
  (HR) compressible flow, radiation diffusion/transport and electromagnetics.}
\label{fig:amr}
\end{figure}

Tensor-product mesh elements (quadrilaterals in 2D and hexahedra in 3D) are
attractive in many high-order applications, because their tensor product structure
enables efficient operator evaluation (see e.g. Section \ref{sec:benchmarks}),
as well as refinement flexibility (e.g. {\em anisotropic} refinement). Unlike
the conforming case however, {\em hanging} nodes that occur after local
refinement of quadrilaterals and hexahedra are not easily avoided by further
refinement. Therefore, CEED researchers are interested in {\em non-conforming}
(irregular) meshes, in which adjacent elements need not share a complete face or
edge and where some finite element degrees of freedom (DOFs) need to be
constrained to obtain a conforming solution.

The MFEM finite element library provides general support for such non-conforming
adaptive mesh refinement, including anisotropic refinement, derefinement and
parallel load balancing. In order to support the entire de Rham sequence of
finite element spaces, at arbitrarily high-order, we use a variational
restriction approach to AMR described in \citep{mfem-amr}. This approach
naturally supports high-order curved meshes, as well as finite element
techniques such as hybridization and static condensation. It is also highly
scalable, easy to incorporate into existing codes, and can be applied to complex
(anisotropic, $n$-irregular) 3D meshes, see Figure \ref{fig:amr}.

The CEED conforming mesh generation capability builds on the PUMI/MeshAdapt
\citep{ibanez2016pumi} libraries developed as part of the FASTMath SciDAC
applied math institute. Within PUMI the curved mesh entities, edges, faces and
regions, are represented as Bezier polynomials \citep{farin2014curves}. The use
of the Bezier properties, curve containment in the convex hull of control
points, derivatives and products of Bezier functions being Bezier functions, and
the existence of efficient degree elevation and subdivision algorithms, simplify
the definition of curved mesh entity operations. One critical operation is the
conversion of curved mesh Bezier geometry into interpolating geometry that is
common input to analysis codes. The MeshAdapt procedures employ cavity based
mesh modification operators that include optimization based entity curving, mesh
entity refinement, mesh cavity coarsening, and mesh cavity swap operations
\citep{lu2014parallel,LuoShephard_04}. The input to MeshAdapt is an anisotropic
mesh metric field defined over the entities of the current mesh. The mesh metric
field can be defined as any combination of sizes as dictated by error
estimation/indication procedures, feature based detection operators or other
user defined size field information. Given a mesh size field MeshAdapt carries
out a series of cavity based operations to modify the local mesh topology and/or
geometry to satisfy the requested mesh size field. The current curved mesh
adaptation procedures operate on CPUs. Efforts have been initiated to extend the
GPU based Omega\_h \citep{ibanezPhd, osh_github} straight edged mesh adaptation
procedures to support curved mesh entities and to include additional mesh
modification operators used in curved mesh adaptation.

Controlling element shapes for evolving meshes when curved elements are used
introduces additional complexity past those encountered when straight edge
elements are used. In particular, methods are needed to effectively support the
definition of well shaped elements in the application of ALE methods in
Lagrangian reference frame simulations when meshes become highly deformed, or in
the application of cavity based curved mesh modifications where new curved mesh
entities must be defined within a curved mesh cavity. Methods that apply direct
curved element shape optimization are being used to address these needs
\citep{dobrev2019target, feuillet2018connectivity}.

\subsection*{Performance Portability}

\begin{figure}[htbp]
  \centering
  \includegraphics[width=0.6\linewidth]{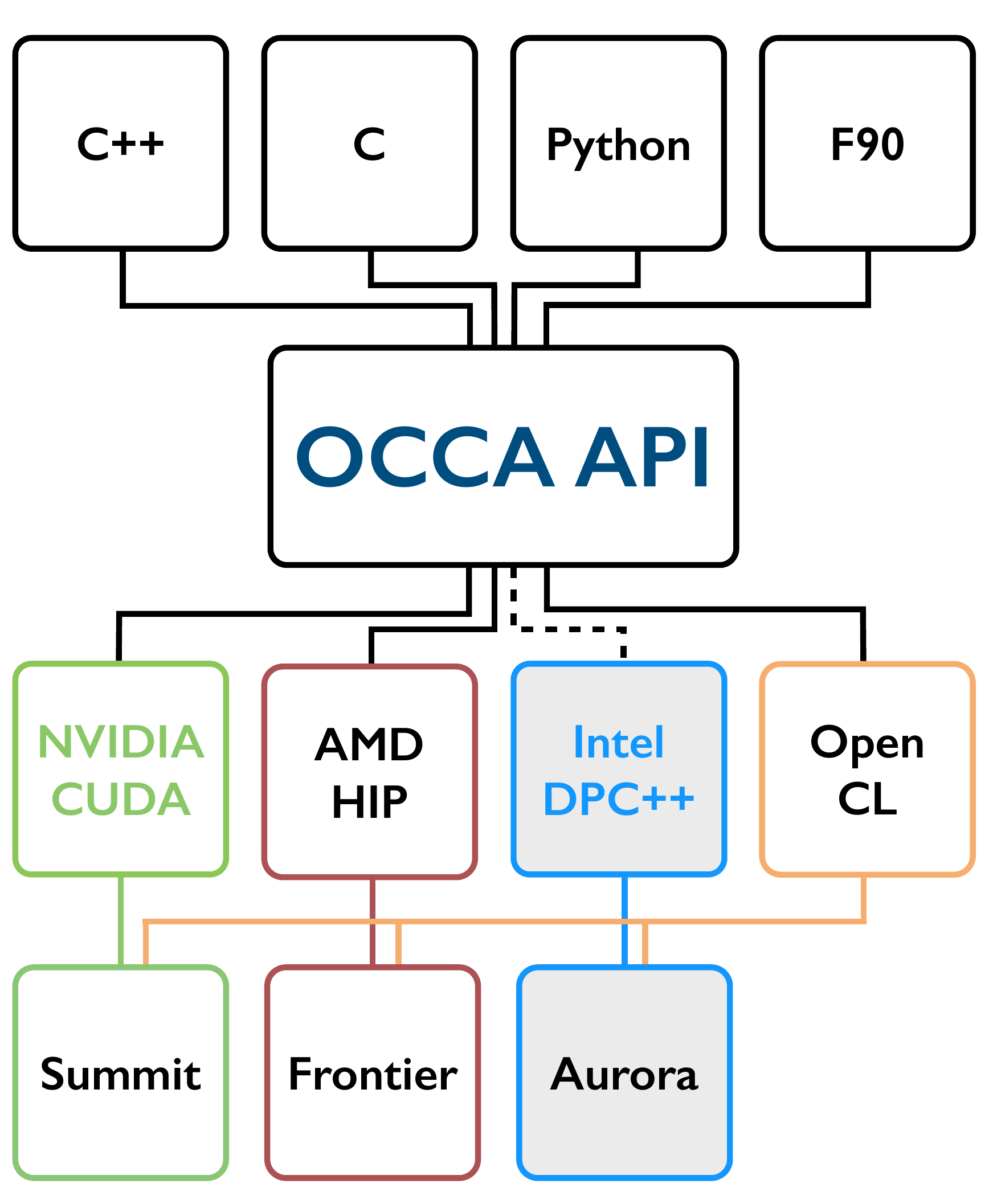}
  \caption{The OCCA portability layer provides a unified API for offloading
    computation to multiple backends. The Intel OneAPI backend is currently in
    progress.}
  \label{occa-diagram}
\end{figure}

\noindent The MFEM, libCEED, NekRS, and libParanumal software packages developed
as part of the CEED project all include support for performance portability
achieved to varying degrees using the Open Concurrent Compute Abstraction (OCCA)
\citep{occa,occa-web}. As pictured in Figure \ref{occa-diagram} OCCA includes
APIs for C, C++, F90, and Python. It provides multiple backends enabling
portability to GPUs programmed using CUDA, OpenCL, and HIP. A new DPC++ OCCA
backend is in development to provide native support for upcoming Intel discrete
GPUs. Several of these programming models also enable cross platform portability
providing additional options to achieve cross platform efficiency.

OCCA exposes all performance critical features of the support backends required
for high-order finite element calculations, enabling performance tuning of
kernels that can achieve performance similar to kernels written to target the
backends directly. We take advantage of the OCCA capability to compile compute
kernels at run-time with just-in-time (JIT) specialization and optimization,
which is particularly important for high-order methods where innermost loops
have bounds depending on the order.

\subsection*{Small Tensor Contractions}

\begin{figure*}
\centering
\includegraphics[width=0.8\textwidth]{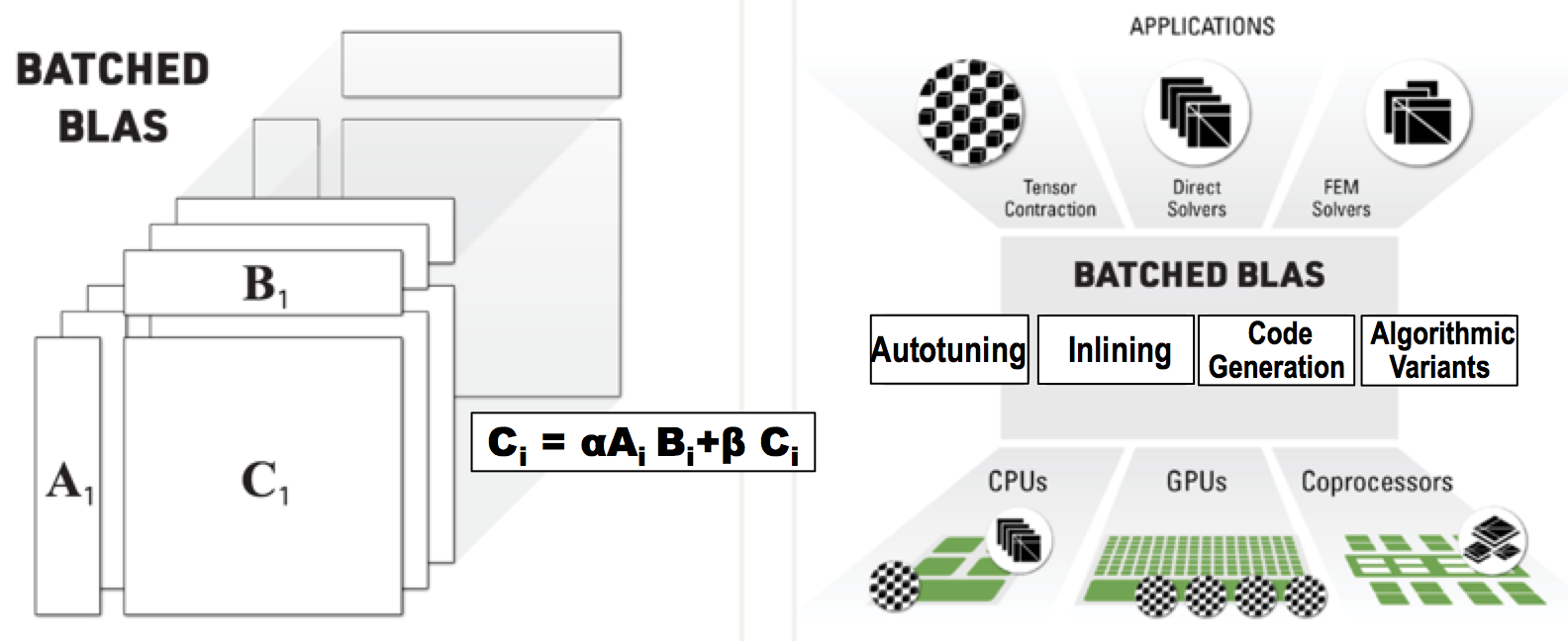}
\caption{Standardizing a Batched BLAS API, an extension to the BLAS standard,
  enables users to perform thousands of small BLAS operations in parallel while
  making efficient use of their hardware.}
\label{fig:bblas}
\end{figure*}

The numerical kernels of efficient high-order operator evaluation reduce to many
small dense tensor contractions, one for each element of the computational
mesh. These contractions can be performed in parallel over the elements and can
be implemented as a batch of small matrix-matrix multiplications
({\code{DGEMMs}}, see Figure~\ref{fig:bblas}). Vendor-optimized BLAS routines
have been successfully used in many areas to provide performance portability
across architectures. Similarly, the availability of highly optimized Batched
BLAS for various architectures can provide tensor contractions, and consequently
high-order applications, performance portability. Therefore, CEED scientists
have been working with vendors and the community on defining a Batched BLAS API,
and finalized a proposed API for Batched BLAS~\citep{batch-api,BBLAS2}.

The MAGMA library provides the most complete set of highly optimized Batched
BLAS, including batched {\code{DGEMMs}} on GPUs. Very small batched
{\code{DGEMMs}} have been optimized to perform at their theoretical performance
upper bounds for a number of
architectures~\citep{abdelfattah2016high,masliah2016high}. Furthermore, the
tensor contraction kernels in CEED often require a sequence of batch
{\code{DGEMMs}}. Such calls can share the same execution context so that they
operate on the fast memory levels of the hardware, thus maximizing the memory
bandwidth~\citep{ceed-ms32}.

In addition, CEED has modes of operation where the elementwise operator
evaluation can be recast as standard batch {\code{DGEMMs}} on
medium-to-large-sized matrices~\citep{abdelfattah2016performance}; the MAGMA
backend for libCEED exploits this to improve performance for non-tensor finite
elements~\citep{ceed-ms34}. The use of the batch BLAS operations increases the
chances of performance portability, since BLAS is often highly optimized by
vendors and other open source numerical software. This was recently illustrated
with the MAGMA port and CEED backend for AMD
GPUs~\citep{hipmagma-hpec,cade_brown_2020_3928667,ceed-ms34}.

\subsection*{Matrix-free Linear Solvers}

In addition to efficient discretization and operator evaluation, matrix-free
preconditioning is essential in order to obtain highly performant solvers at
high order. Solvers based on explicitly formed matrices tend to have low
arithmetic intensity, and the memory requirements associated with the system
matrices for high-order discretizations are typically too large to be practical
on GPUs and accelerator-based architectures. On the other hand, many standard
preconditioning techniques rely on the knowledge of the matrix entries. For
these reasons, matrix-free preconditioning is both an important and challenging
topic.

Multigrid methods provide one promising avenue for the development of matrix
free linear solvers \citep{Kronbichler2019}. These methods have optimal
complexity, and when combined with effective matrix-free smoothers, have the
potential to achieve excellent performance \citep{Lottes2005}. Recent work has
also studied the matrix-free construction of fast diagonalization smoothers for
discontinuous Galerkin methods \citep{Pazner2018e}. Both $h$-multigrid, where a
sequence of geometrically coarsened meshes is used, and $p$-multigrid, in which
a hierarchy of polynomial degrees is constructed, can be used in conjunction to
obtain an efficient solver \citep{Sundar2015}. At the coarsest level, algebraic
multigrid (AMG) methods, such as those from the \textit{hypre} software library,
are required in order to obtain a truly scalable solver.

An additional technique used to precondition high-order systems is to assemble a
spectrally equivalent sparsified system, to which standard matrix-based
preconditioning techniques may be applied. One method of obtaining a spectrally
equivalent sparse matrix is using a low-order discretization on a refined mesh,
and making use of the so-called finite element method--spectral element method
(FEM--SEM) equivalence for tensor-product elements
\citep{sao80,Canuto1994,Canuto2006}. Recent work has demonstrated that, when
combined with efficient solvers for the sparsified system, this approach can
result in highly efficient solvers \citep{Bello-Maldonado2019,Pazner2019a}. One
challenging property of the resulting low-order refined system is that the
meshes resulting from the refinement procedure are not shape regular with
respect to the polynomial degree $p$: the aspect ratio of the mesh elements
increases with increasing polynomial degree. As a result, algebraic multigrid
methods with pointwise smoothers such as point Jacobi result in degraded
convergence at high orders. Consequently, the development of specialized
matrix-free smoothers for these anisotropic low-order systems is also of
interest. Additionally, the extension of these low-order preconditioners to
high-order simplex elements is a topic of ongoing research \citep{Chalmers2018}.

Also of interest is the development of efficient matrix-free solvers for
$H(\mathrm{curl})$, $H(\mathrm{div})$, and discontinuous Galerkin finite element
spaces. It is often the case that efficient solvers for $H^1$ discretizations
can be modified or supplemented to obtain good preconditioners for these more
challenging cases. For example, multigrid solvers for diffusion problems can be
combined with a discrete gradient operator to obtain uniform preconditioners for
definite Maxwell problems discretized using $H(\mathrm{curl})$ finite elements
\citep{Kolev2009}. Although these solvers were originally developed in the
context of matrix-based AMG, the same ideas can be extended to the matrix-free
setting. Furthermore, uniform preconditioners for $H^1$ conforming diffusion
problems can be combined with a simple diagonal scaling to obtain uniform
preconditioners for DG diffusion problems \citep{Antonietti2016,Dobrev2006}.

An additional method that is capable of using fast diagonalization methods (for
operators that admit separable approximations) is Balancing Domain Decomposition
by Constraints (BDDC) \citep{dohrmann2003psb,zampini2016pcbddc}, which offers
more localized smoother construction, faster convergence for additive cycles,
and more rapid coarsening than the fast diagonalization technique discussed
above. BDDC has been used for high order elements applied to almost
incompressible elasticity \citep{pavarino2010bddc}, where the condition number
of the BDDC-preconditioned operator for single-element smoothing and coarsening
was shown to scale as $ \kappa \le C \big(1 + \log p^2 \big)^2, $ where $p$ is
the polynomial degree and $C$ is robust to element size/shape and the Poisson
ratio. BDDC has also been analyzed as a multigrid method
\citep{brown2019lfabddc}, and can be composed with other multigrid methods.

\subsection*{High-order Data Analysis and Visualization}

Accurate visualization of general finite element meshes and functions in the de
Rham complex requires finite element knowledge that may not be present in
visualization tools employed by applications. The visualization needs to account
for the orders of the mesh and solution fields, as well as the type of finite
element basis used for each of them. Our work in this direction is based on the
current capabilities in MFEM, illustrated in its native GLVis visualization tool
\citep{glvis}, as well as in the VisIt visualization and data analysis
application \citep{VisIt_web}.

An additional challenge for high-order meshes and functions is that there is no
common community standard for the description of high-order data at arbitrary
other. CEED is working with visualization and application teams to develop a
standard called Field and Mesh Specification (FMS) that not only improves
visualization capabilities but also enables consistent data transfer between
high-order applications. See \citep{fms} and \citep{ceed-ms18}.

%% file: tex/benchmarks.tex
\section{Benchmarks} \label{sec:benchmarks}

\begin{figure*}[t] \centering
 \includegraphics[width=0.335\textwidth]{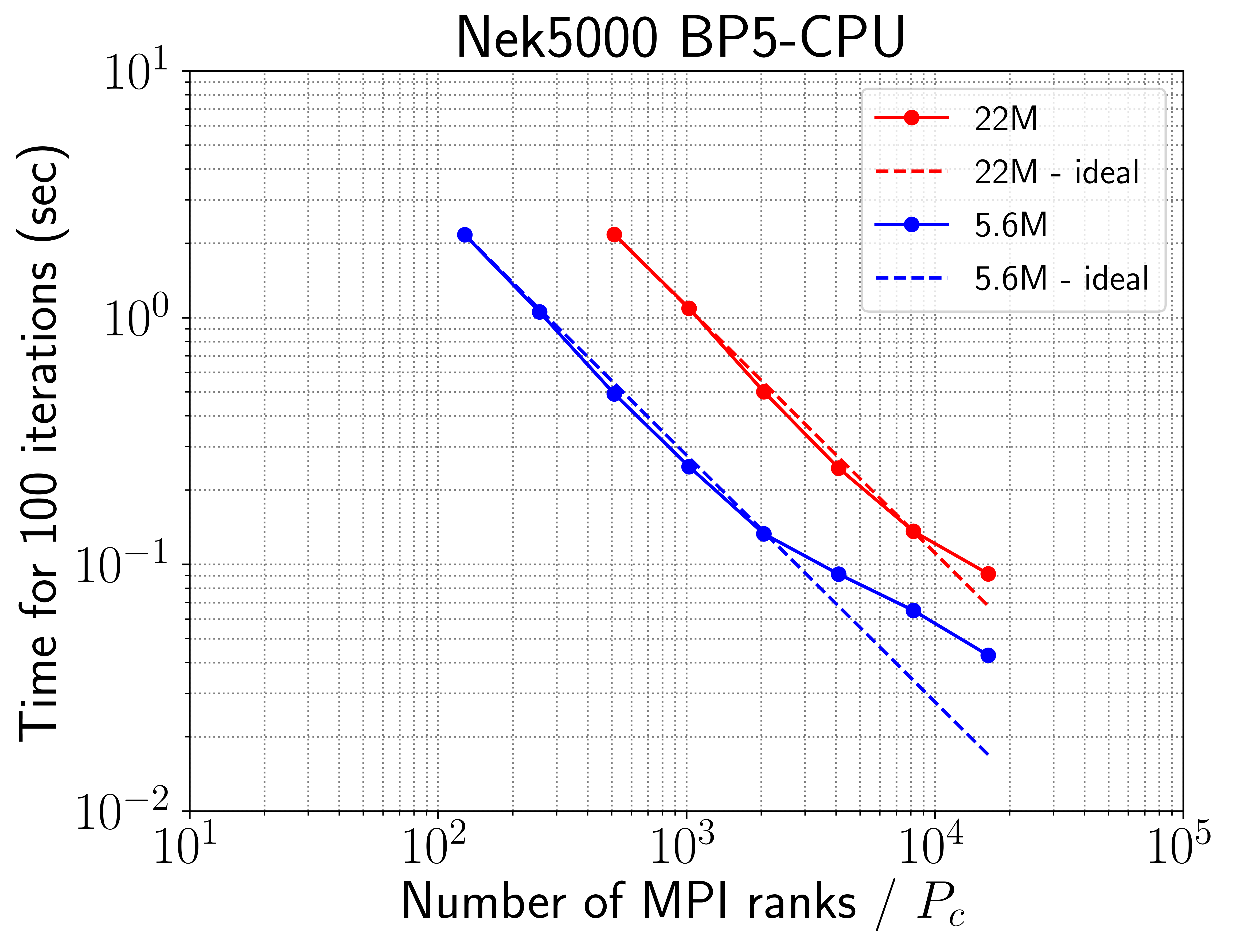}
 \includegraphics[width=0.335\textwidth]{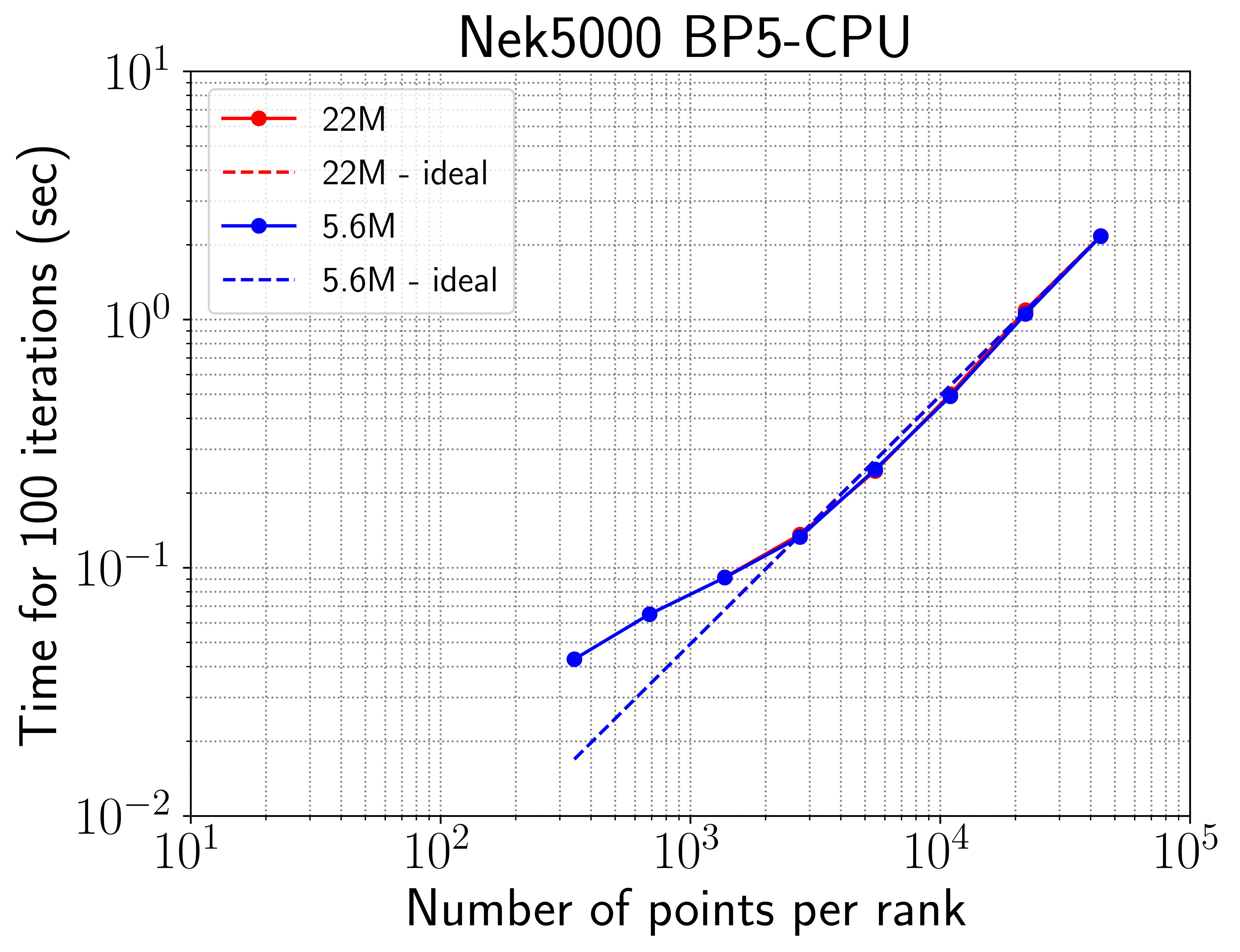}
 \includegraphics[width=0.315\textwidth]{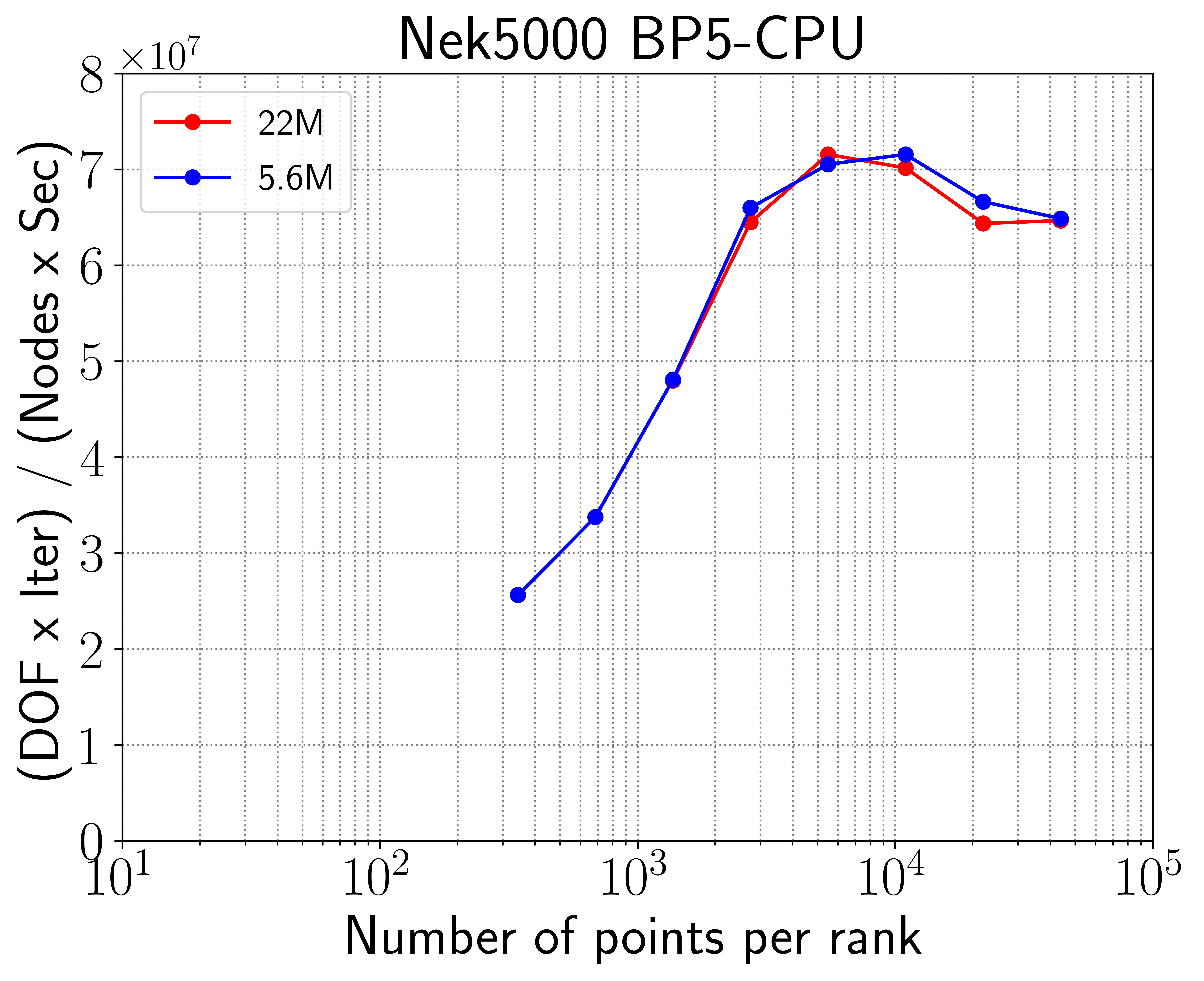}
\caption{\label{fig:nek5_mira}Strong-scaling analysis for BP5 in -c32 mode on
  Mira: (left) standard strong-scaling plots with increasing number of MPI
  ranks, $P_c$ for Poisson problems using $n$=5.6M and 22M grid points; (center)
  data collapse manifest when independent variable is $n/P_c$; and (right)
  work-rate (left ordinate) and parallel efficiency (right ordinate) vs
  $n/P_c$.}
\end{figure*}

Application-relevant performance testing and analyses are critical to effective
HPC software deployment. One of the foundational components of CEED is a
sequence of PDE-motivated bake-off problems (BPs) designed to establish best
practices for performant implementations of high-order methods across a variety
of platforms. The idea is to pool the efforts of multiple high-order development
groups to identify effective code optimization strategies for candidate
architectures. In an initial round of tests we compared performance from four
software development projects (Nek5000, MFEM, deal.II, and libParanumal) on
Mira, the BG/Q at ALCF, and Summit, the NVIDIA V100-based platform at ORNL. The
results of this bake-off were documented in \citep{ceed_bp_paper_2020}. We are
interested in peak performance (degrees of freedom per second, per node) and in
strong-scale performance at a significant fraction of this peak (e.g., 80\%), as
this regime is frequently of paramount concern to computational scientists.
While we consider matrix-free implementations of $p$-type finite and spectral
element methods as the principal vehicle for our study, the performance results
are relevant to a broad spectrum of numerical PDE solvers, including finite
difference, finite volume, and $h$-type finite elements, and thus are widely
applicable.

The first suite of CEED bake-off problems, BP1--BP6, is focused on simple solver
kernels---conjugate gradient (CG) iterations to solve systems of the form
$
\left( \alpha A \;+\; \beta B \right) \uu_i = B \uf_i,
$
which are the discrete equivalents of the constant-coefficient 3D
positive-definite Helmholtz problem,
\begin{eqnarray} \nonumber
-\alpha \nabla^2 u_i \;+\; \beta u_i &=& f_i(\bx), \;\;\; \mbox \bx \in \Omega \subset \RR^3,
\end{eqnarray}
for $i=1,\dots,m$, with homogeneous Dirichlet conditions, $u_i=0$ on $\dO$. The
odd-numbered BPs correspond to scalar problems ($m=1$), whereas the
even-numbered cases correspond to (potentially more efficient) vector problems
($m=3$). An important aspect of using CG is that it involves a mix of local work with both
nearest neighbor and global communication (vector reductions), which provides at
least moderate stress on the system communication.

The BP discretizations are based on isoparametric $Q_p$ finite elements
(curvilinear bricks) on a tensor product reference domain, $\br \in \Oh
=[-1,1]^3$, which is mapped through a transformation $\bx^e(\br)$ for each of
$E$ elements, $\Omega^e$, $e=1,\dots,E$. Denoting the underlying
$C^0$-Lagrangian basis functions as $\phi_i(\bx)$, $i=1,\dots,n$, the respective
stiffness and mass matrix entries are
\begin{eqnarray*}
  A_{ij} = \int_{\Omega} \nabla \phi_i \cdot \nabla \phi_j \,dV,
  &\hspace{.1in}&
  B_{ij} = \int_{\Omega} \phi_i \phi_j \,dV.
\end{eqnarray*}
These matrices are never formed, but instead are applied using fast,
low-storage, tensor-product-sum factorization that are at the heart of efficient
high-order methods \citep{dfm02,sao80}.

\begin{figure*}
\centering
\includegraphics[width=0.315\textwidth]{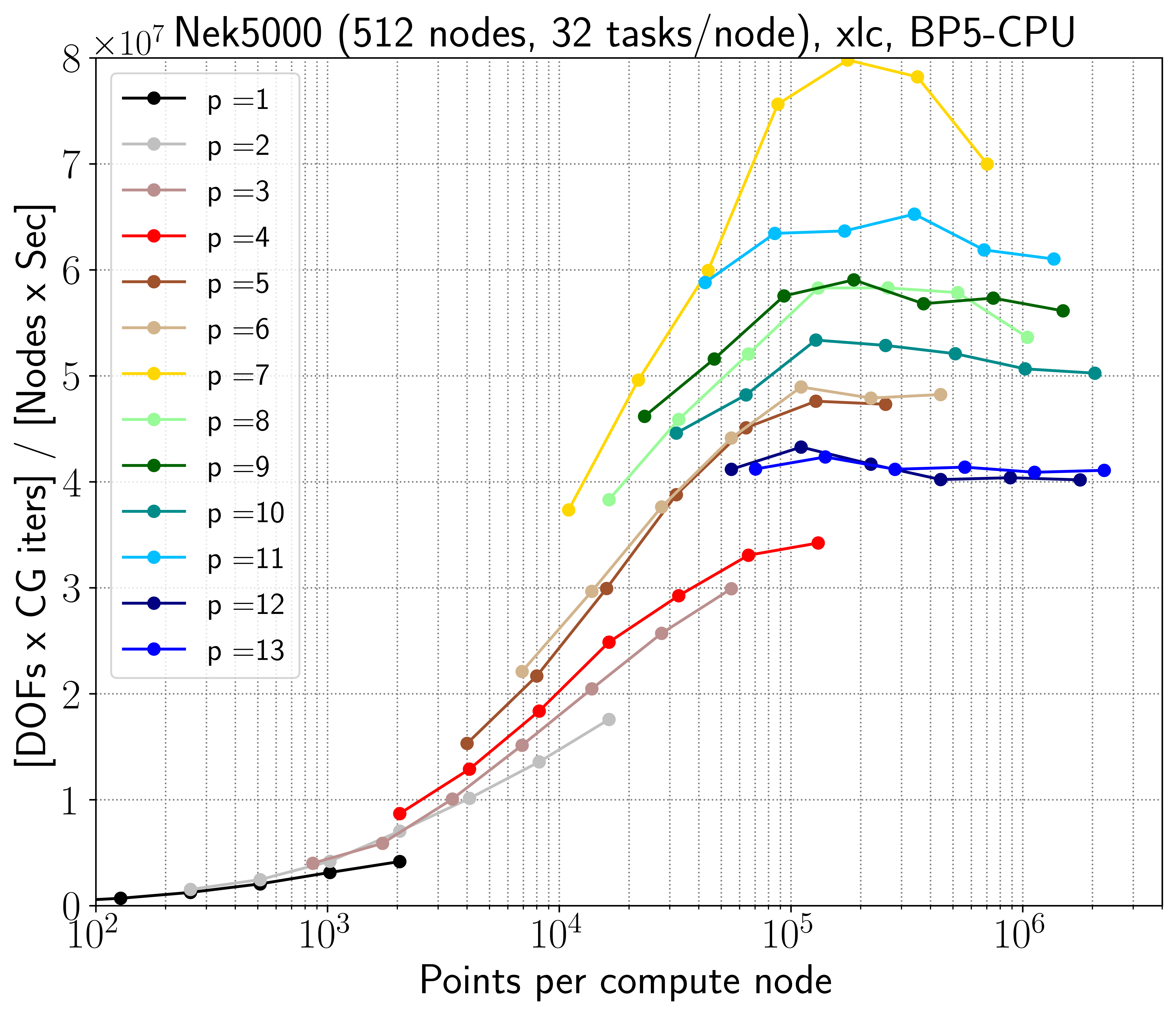}
\includegraphics[width=0.33\textwidth]{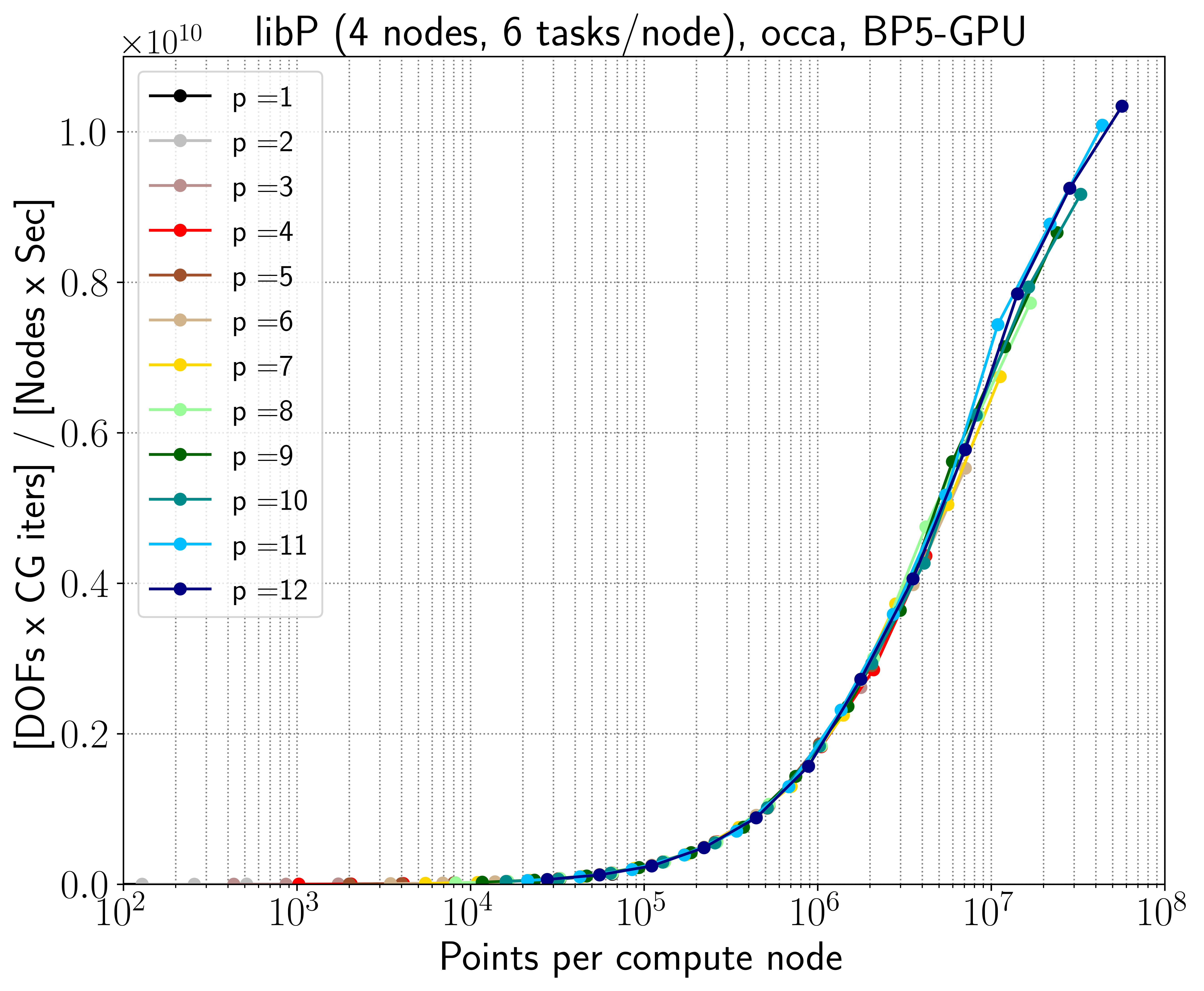}
\includegraphics[width=0.33\textwidth]{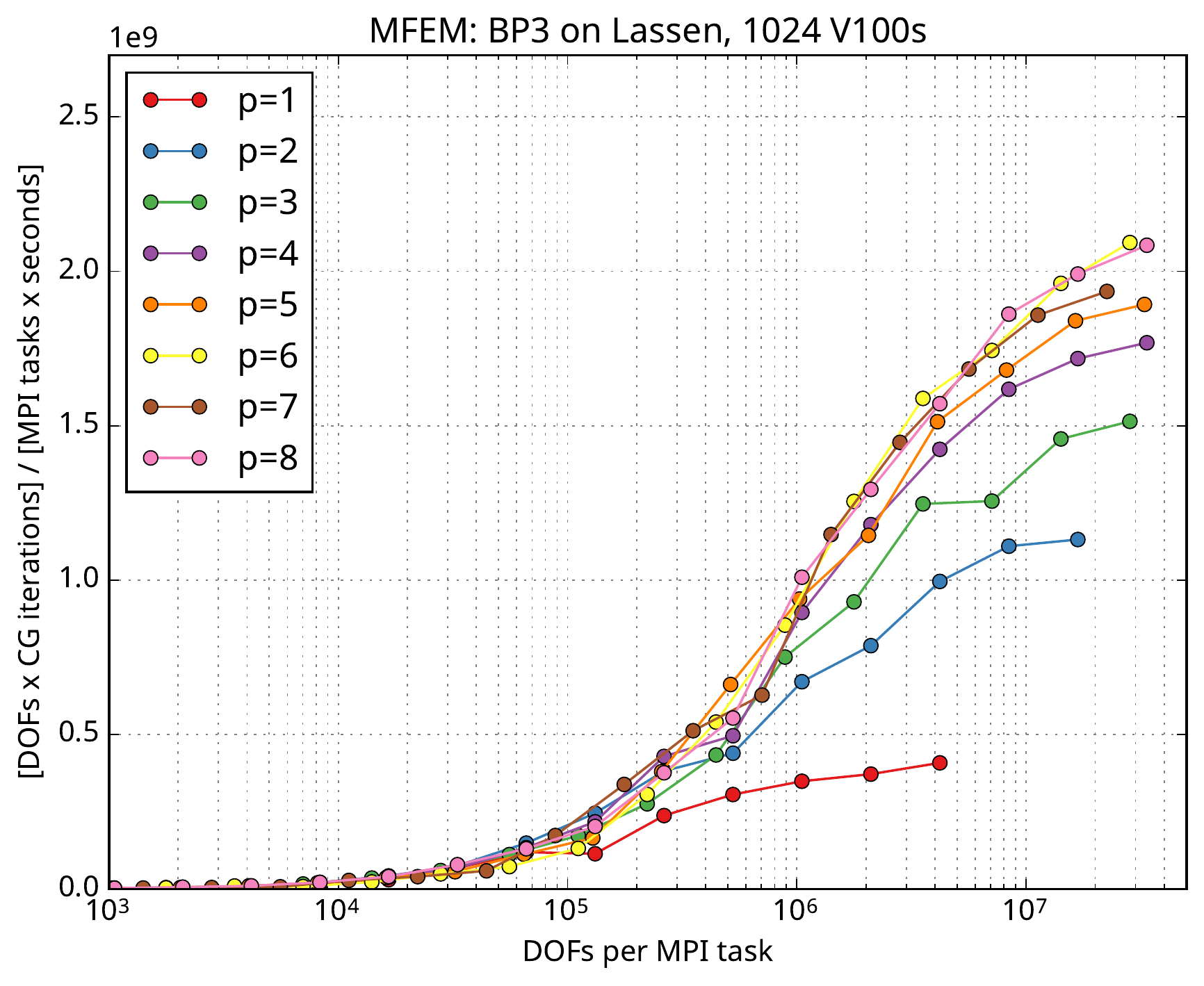}
\caption{\label{fig:bps_ecp}
BP results:
BP5 with Nek5000 on Mira;
BP5 with libParanumal on Summit;
BP3 with MFEM on Lassen
}
\end{figure*}

Test problems BP1--BP2 correspond to solving the mass matrix ($\alpha=0$,
$\beta=1$), while BP3--BP6 correspond to solving the Poisson problem
($\alpha=1$, $\beta=0$). For BP1--BP4, integration is performed over each
element using Gauss-Legendre quadrature with $q=p+2$ nodes in each direction in
$\Oh$. BP5--BP6 correspond to the spectral element formulation, in which
integration is performed on the underlying $(p+1)^3$ Gauss-Lobatto-Legendre
nodal points, thus bypassing interpolation from nodes to quadrature points.

An important question in the development of HPC software is to ensure that
testing reflects actual use modalities.
On large HPC platforms, users typically use as many nodes as are effective,
meaning that they run at the {\em strong-scale limit}, rather than the
work-saturated limit.
Figure \ref{fig:nek5_mira} illustrates these limits for the case of BP5 on up to
16,384 MPI ranks on Mira. On the left we see standard strong-scale plots for two
different problem sizes, $n=5.6$ million points and $n=22$ million points. The
smaller case exhibits perfect linear speed-up up to $P_c=2048$ MPI ranks whereas
the larger case sustains linear speed-up out to $P_c=8192$ ranks. For this class
of problems with a given code and platform the dominant factor governing
parallel efficiency is the number of points per node (or core, or other
independent compute resource) \citep{fischer15}. Indeed, with this metric we see
a perfect data collapse in Figure \ref{fig:nek5_mira} (center), which shows the
time as a function of the number of points per rank, and (right), which shows
the work-rate (DOFS=degrees-of-freedom $\times$ number of iteration per second
per node) and the parallel efficiency, $\eta = T_1 /(P T_{P})$, where $T_P$ is
the time when running on $P$ MPI ranks.

We make several observations about Figure \ref{fig:nek5_mira} (right). First,
the strong-scale limit is at about 2700 points per rank. Running with more
points per rank keeps the efficiency at unity but increases the runtime. Running
with fewer points per rank means increasing the total number of cycles
(core-hours) to complete the job. Very often, users will trade some degree of
inefficiency for decreased runtime. If we choose, for example, 80\% efficiency, the
value of $n/P$ where this value is realized is denoted by $n_{0.8}$. Second, it
is beneficial to increase rate of work (DOFS) because fewer core-hours are then
required to complete the overall task. Wall-clock time, however, may not be
reduced if increasing the work-rate implies an increase in $n/P$ to stay above
the token (e.g., 80\%) efficiency mark. For a problem of size $n$, the
time-to-solution will be $t_{\eta} = C \frac{n} {\eta \cdot P\cdot r_{\max}}$,
where $\eta$ is the parallel efficiency, $r_{\max}$ is the saturated work rate
(e.g. per computational node), $P$ is the number of nodes used, and $C$ is a
constant that reflects the amount of work per gridpoint. The choice $\eta=0.8$
implies that $n/P=n_{0.8}$, such that the run time is $t_{0.8} = \frac{C}{0.8}
\frac{n_{0.8}}{r_{\max}}$. The problem size $n$ and number of nodes drop out of
the run time formula---the only thing that influences {\em time-to-solution} at
the strong-scale limit is the {\em ratio} of the local problem size to peak
processing rate, $n_{0.8}/r_{\max}$. Minimization of this ratio is of paramount
importance for reduced run time.

Motivated by the preceding analysis, we routinely collect performance data for
BP1--BP6 for varying problem sizes $n=Ep^3$ on a variety of platforms. A
comprehensive study entailing more than 2000 trials was reported in
\citep{ceed_bp_paper_2020}, which considered $E=2^{14}$ to $E=2^{21}$ and $p=1$ to $p=16$
using MFEM, Nek5000, and deal.ii on $P=512$ nodes on ALCF's Mira (in -c32 mode)
and the OCCA-based libParanumal code with $P=4$ (24 NVIDIA V100s) on OLCF's
Summit. Typical BP results are shown in Figure \ref{fig:bps_ecp}. The left panel
shows BP5 results for Nek5000 on Mira. We see that higher polynomials generally
realize a higher DOFS rate and that $n_{0.8} \approx 50,000$, with $r_{0.8}
\approx 65$ MDOFS. On Summit, libParanumal realizes a peak of more than $10$ GDOFS with
$n_{0.8} \approx 10^7$. We also show recent results for BP3 using MFEM on 256
nodes of the V100-based platform, Lassen. Again, higher polynomial orders
sustain higher DOFS, peaking at $\approx 2$ GDOFS per GPU (8 GDOFS/node) with an
$n_{0.8} \approx 3$ million. The corresponding $t_{0.8}$ for these cases are
approximately .0008s on Mira, .001s on Summit, and .0015s on Lassen. We
reiterate that the strong-scale limit is the fastest point where users can (and
will) run while sustaining their desired efficiency. Thus, performance at other
points on Figures \ref{fig:nek5_mira} and \ref{fig:bps_ecp} are of far less
importance. Performance tuning must focus on moving up and to the left in these
plots.

Future BPs will look at dealiased advection kernels, which are compute and
memory intensive, and optimal preconditioning strategies for high-order
discretizations of elliptic problems that represent computational bottlenecks in
several of the ECP applications.

%% file: tex/miniapps.tex
\section{Miniapps} \label{sec:miniapps}

CEED is developing a variety of miniapps encapsulating key physics and numerical
kernels of high-order applications. The miniapps are designed to be used in a
variety of co-design activities with ECP vendors, software technologies projects
and external partners. For example, several of the CEED miniapps (Nekbone and
Laghos) are used as vendor benchmarks in the DOE's CORAL-2 and LLNL's CTS-2
procurements.

\subsection*{libParanumal}

\begin{figure}[htbp]
  \centering
  \includegraphics[width=0.8\linewidth]{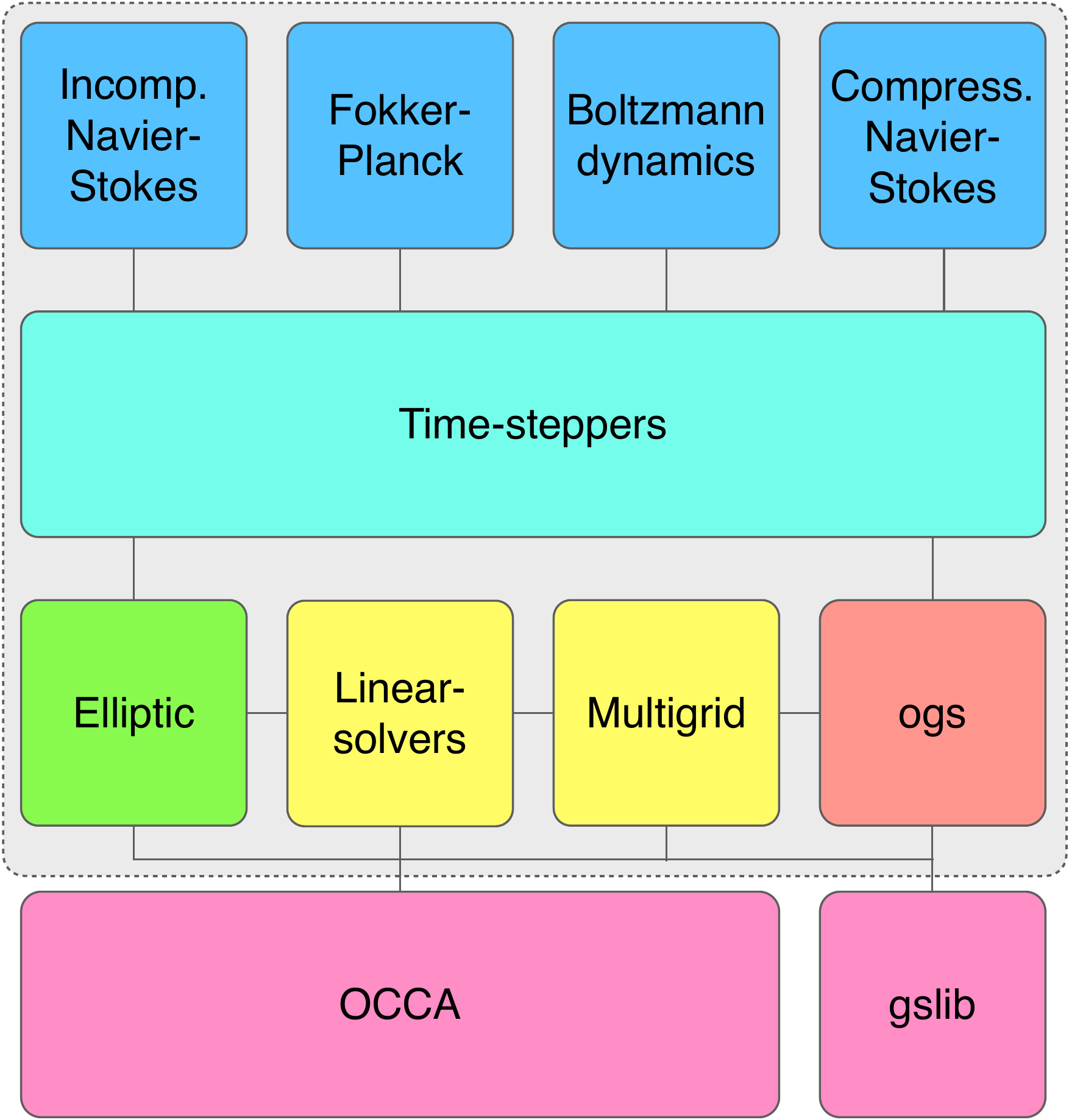}
  \caption{The libParanumal library includes portable GPU accelerated miniapps
    for solving the incompressible and compressible Navier-Stokes,
    Fokker-Planck, and finite moment Boltzmann gas dynamics equations among
    others.}
  \label{libp-diagram}
\end{figure}

libParanumal (LIBrary of PARAllel NUMerical ALgorithms) is an open source
project \citep{ChalmersKarakusAustinSwirydowiczWarburton2020} under development
at Virginia Tech. It consists of a collection of miniapps with high-performance
portable implementations of high-order finite-element discretizations for a
range of different fluid flow models. The miniapps embedded in libParanumal
include solvers for incompressible flows \citep{karakus2019gpu}, compressible
flows, finite moment Boltzmann gas dynamics models
\citep{karakus2019discontinuous}, and Fokker-Planck models. All of these
miniapps are accompanied with highly performant GPU kernels for high-order
Galerkin \citep{swirydowicz2019acceleration} and/or discontinuous Galerkin
spatial discretizations with a collection of high-order time integrators.
libParanumal is constructed as a set of core libraries as shown in Figure
\ref{libp-diagram} including high-performance scalable preconditioned iterative
Krylov subspace solvers with optional multigrid preconditioning. All
computationally intensive calculations are implemented using kernels compatible
with the OCCA portability layer \citep{occa} and have been analyzed and
optimized to guarantee that they achieve a high percentage of the attainable
DEVICE memory bandwidth on NVIDIA P100 \citep{swirydowicz2019acceleration} and
V100 GPUs \citep{ceed_bp_paper_2020}. The libParanumal library provides core GPU
acceleration capabilities to NekRS and algorithms developed for it have been
deployed in the MFEM cuda-gen backend.

\subsection*{NekBench and Nekbone}

\begin{figure*}[t]
\begin{center}
\includegraphics[width=0.98\textwidth]{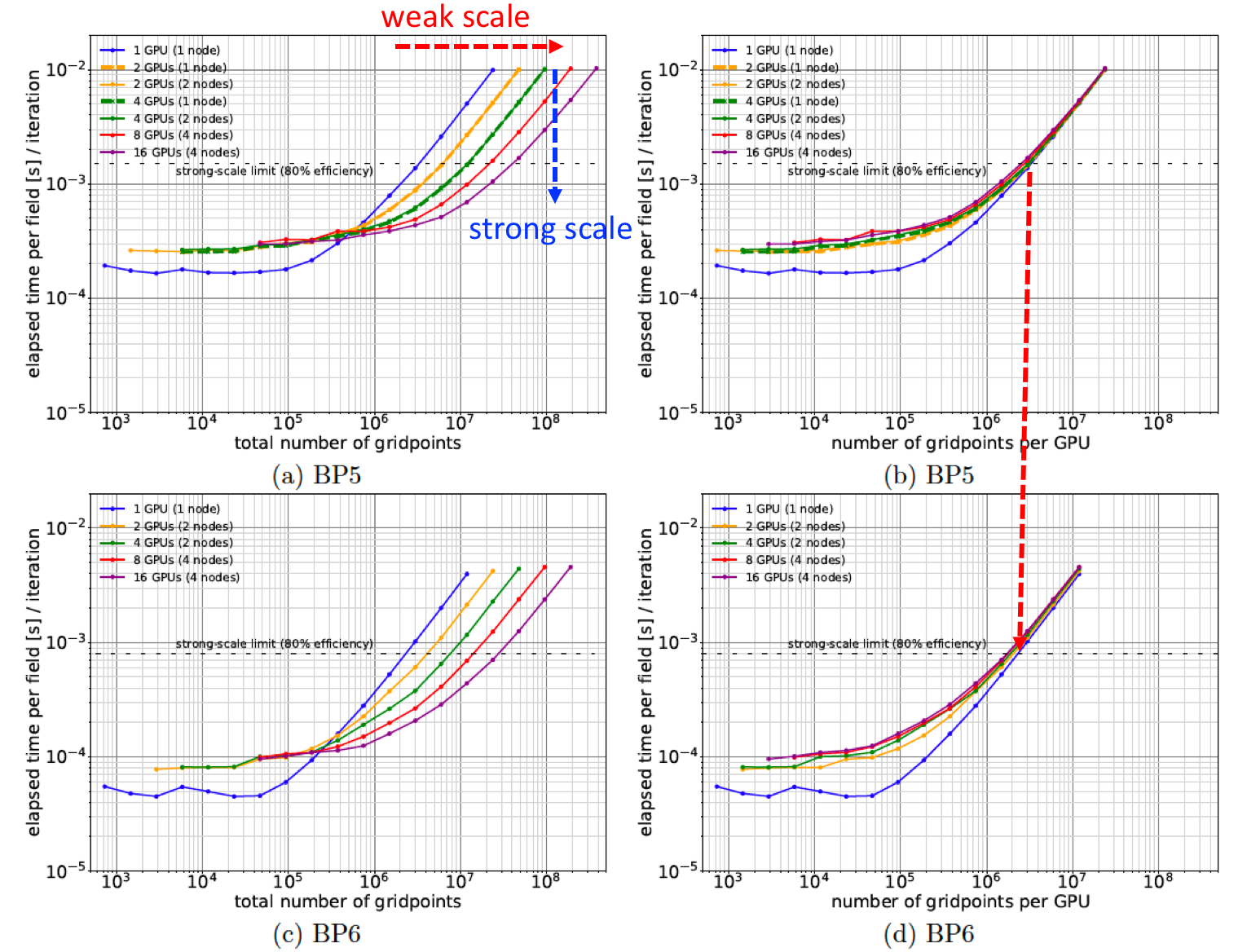}
\caption{
Strong and weak scaling studies of BP5 and BP6 using V100s on Lassen.}
\label{nek-miniapp}
\end{center}
\end{figure*}

NekBench is a benchmark suite representing key components of Nek5000/CEM/RS.
This miniapp supports a variety of benchmarks for fundamental analysis in
different architectures. It supports a single-step driver that delivers timing
measurements on CPUs and GPUs for ping-pong (one-to-one and bisection-bandwidth
tests), gather-scatter, all-reduce, dot-product, and device-to/from-host
memcopy.  It also performs weak and strong scaling tests for
BK5--BK6\footnote{The BKs are {\em bake-off kernels} that involve only the
local, elementwise, portion of the matrix-vector  products, $A\uu$ or $B\uu$
associated with the corresponding BPs described in Section 4.} and BP5--BP6, as
shown in Figure~\ref{nek-miniapp}. The figure demonstrates that $n_{0.8}$ is
reduced from 3M points to 2M points per V100 when we switch from
the scalar (BP5) solver to the vector (BP6) variant of the solver.

Nekbone solves a standard Poisson equation using conjugate gradient iteration
with a simple diagonal preconditioner on a block or linear geometry. It
encapsulates one of the principal computational kernels pertinent to Nek5000,
which includes a mixture of local (near-neighbor) and nonlocal (vector reduction)
communication patterns that are central to efficient multilevel solvers.
Nekbone has been updated to include vector solutions, which allows amortization of
message and memory latencies. Nekbone has been used for assessment of advanced
architectures and for evaluation of light-weight MPI implementations on the ALCF
BG/Q, Cetus, in collaboration with Argonne's MPICH team \citep{raffenetti17}.

\subsection*{Laghos and Remhos}

Laghos (LAGrangian High-Order Solver) and Remhos (REMap High-Order Solver) are
MFEM-based miniapps developed by the CEED team. The objective of these miniapps
is to provide open source implementations of efficient discretizations for
Lagrangian shock hydrodynamics (Laghos) and field remap (Remhos) based on
high-order finite elements.

Laghos \citep{laghos} solves the time-dependent Euler equations of compressible
gas dynamics in a moving Lagrangian frame. The miniapp is based on the method
described in \citep{Dobrev2012}. It exposes the principal computational kernels
of explicit time-dependent shock-capturing compressible flow, including the
FLOP-intensive definition of artificial viscosity at quadrature points.

Laghos supports two options for deriving and solving its system of equations,
namely, the full assembly and the partial assembly methods. Full assembly relies
on global mass matrices in CSR format; this option is appropriate for first or
second order methods. Partial assembly utilizes the tensor structure of the
finite element spaces, resulting in less data storage, memory transfers and
FLOPs; this option is of interest in terms of efficiency for high-order
discretizations. The Laghos implementation includes support for hardware
devices, such as GPUs, and programming models, such as CUDA, OCCA, RAJA and
OpenMP, based on MFEM 4.1 or later. These device backends are selectable at
runtime. Laghos also contains an AMR version demonstrating the use of dynamic
adaptive mesh refinement for a moving mesh with MFEM.

\begin{figure*}[!ht]
\centering
\includegraphics[width=0.48\textwidth]{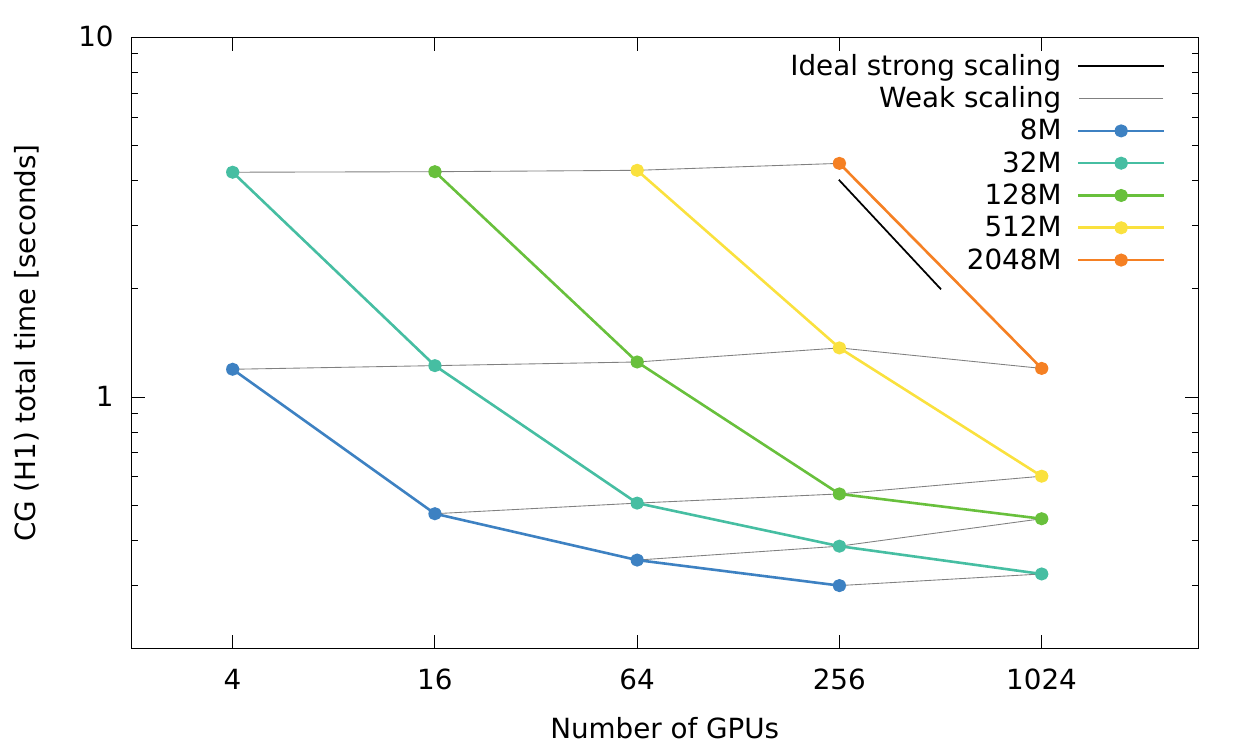}
\includegraphics[width=0.48\textwidth]{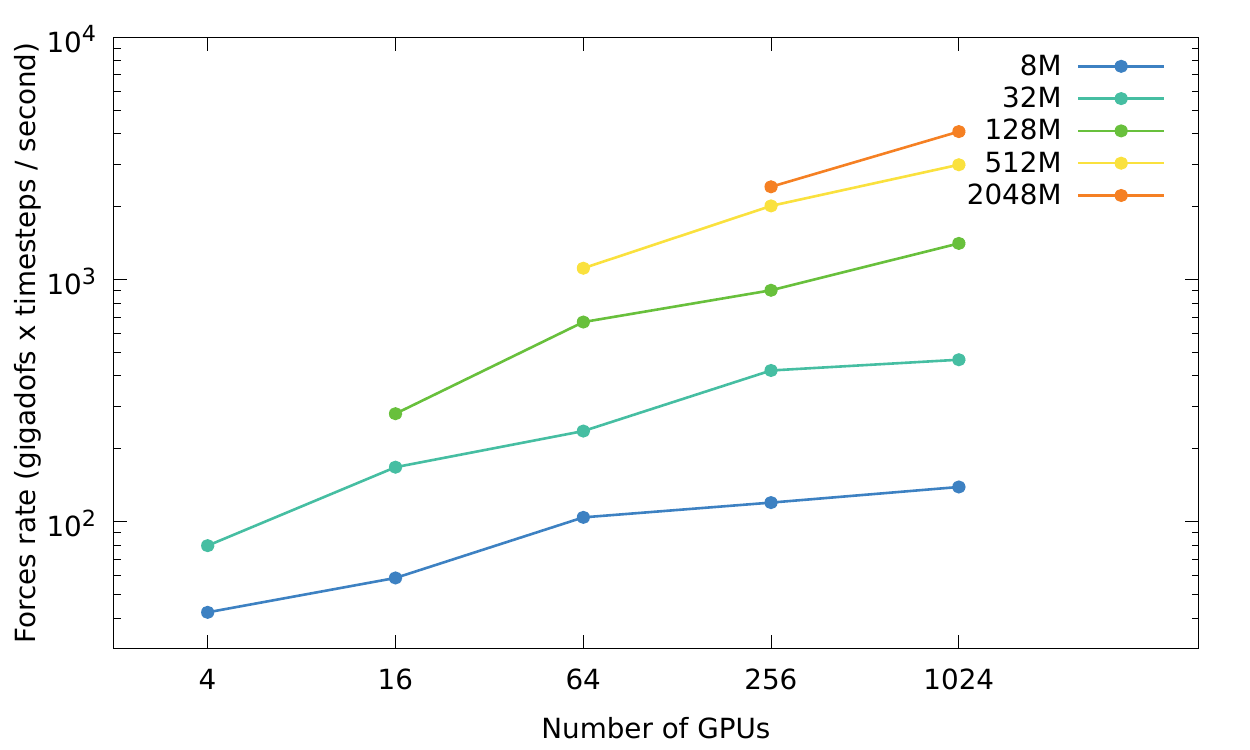}
\caption{Left: weak and strong scaling results with Laghos and MFEM-4.1: 2D
  problem on Lassen, using up to 1024 GPUs. Right: throughput for the
  Laghos force kernel in (GDOF x timesteps / second), reaching above 4 TDOF/s on
  1024 GPUs.}
\label{fig:laghos_time}
\end{figure*}

Large-scale GPU runs of Laghos were performed on Lassen. All computations were
kept on the device, except for the result of the dot product which is brought
back on to the CPU during the iterations of the CG solver. Initial results are
presented in Figure \ref{fig:laghos_time}, showing both the weak (gray lines)
and strong (colored lines) scaling obtained on four to one thousands of GPUs
during the CG iterations of the velocity solver, which corresponds to the
BP2 CEED benchmark. Ideal strong scaling is possible for problem size large
enough, while weak scaling is more easily reached through all the range of the
runs. The bottom panel of Figure \ref{fig:laghos_time} presents the throughput
in DOFs per second for the Laghos force kernel, reaching more than 4 TDOF/s on
the same configuration.

Remhos solves the pure advection equations that are used to
perform conservative and monotonic DG advection-based discontinuous field
interpolation, or ``remap'' \citep{remhos}. Remhos combines discretization methods described in the
following articles: \citep{Anderson2015, Anderson2016, Anderson2018, Hajduk2020,
Kuzmin2020}. It exposes the principal computational kernels of explicit
time-dependent Discontinuous Galerkin advection methods, including monotonicity
treatment computations that are characteristic to FCT (Flux Corrected Transport)
methods.

Remhos supports two execution modes, namely, transport and remap, which result
in slightly different algebraic operators. In the case of remap, the finite
element mass and advection matrices change in time, while they are constant for
the transport case. Just like Laghos, Remhos supports full assembly and partial
assembly options for deriving and solving its linear system. Support for
different hardware devices in Remhos is work in progress.

Other computational motifs supported by both Laghos and Remhos include:
domain-decomposed MPI parallelism; support for unstructured 2D and 3D meshes,
with quadrilateral and hexahedral elements; moving high-order curved meshes;
explicit high-order time integration methods; optional in-situ visualization
with GLVis and data output for visualization and data analysis with VisIt.

%% file: tex/libceed.tex
\section{libCEED} \label{sec:libceed}

libCEED is CEED's low-level API library that provides portable and performant
evaluation of high-order operators \citep{libceed-user-manual}. It is a C99 library
with no required dependencies, and with Fortran and Python interfaces (see for
details on the Python interface \citep{libceed-paper-proc-scipy-2020}).

One of the challenges with high-order methods is that a global sparse matrix is
no longer a good representation of a high-order linear operator, both with
respect to the FLOPs needed for its evaluation, as well as the memory transfer
needed for a matvec. Thus, high-order methods require a new ``format'' that
still represents a linear (or more generally non-linear) operator, but not
through a sparse matrix.

The goal of libCEED is to propose such a format, as well as supporting
implementations and data structures, that enable efficient operator evaluation
on a variety of computational device types (CPUs, GPUs, etc.). This new operator
description, outlined below and in \citep{libceed-user-manual},
is based on algebraically factored form, which is easy to
incorporate in a wide variety of applications, without significant refactoring
of their own discretization infrastructure.

\subsection*{Finite Element Operator Decomposition}

Finite element operators are typically defined through weak formulations of
partial differential equations that involve integration over a computational
mesh. The required integrals are computed by splitting them as a sum over the
mesh elements, mapping each element to a simple \textit{reference} element
(e.g. the unit square) and applying a quadrature rule in reference space.

This sequence of operations highlights an inherent hierarchical structure
present in all finite element operators where the evaluation starts on
\textit{global (trial) degrees of freedom (DOFs) or nodes on the whole mesh},
restricts to \textit{DOFs on subdomains} (groups of elements), then moves to
independent \textit{DOFs on each element}, transitions to independent
\textit{quadrature points} in reference space, performs the integration, and
then goes back in reverse order to global (test) degrees of freedom on the whole
mesh.

\begin{figure*}[ht!]
\centering
\includegraphics[width=0.8\textwidth]{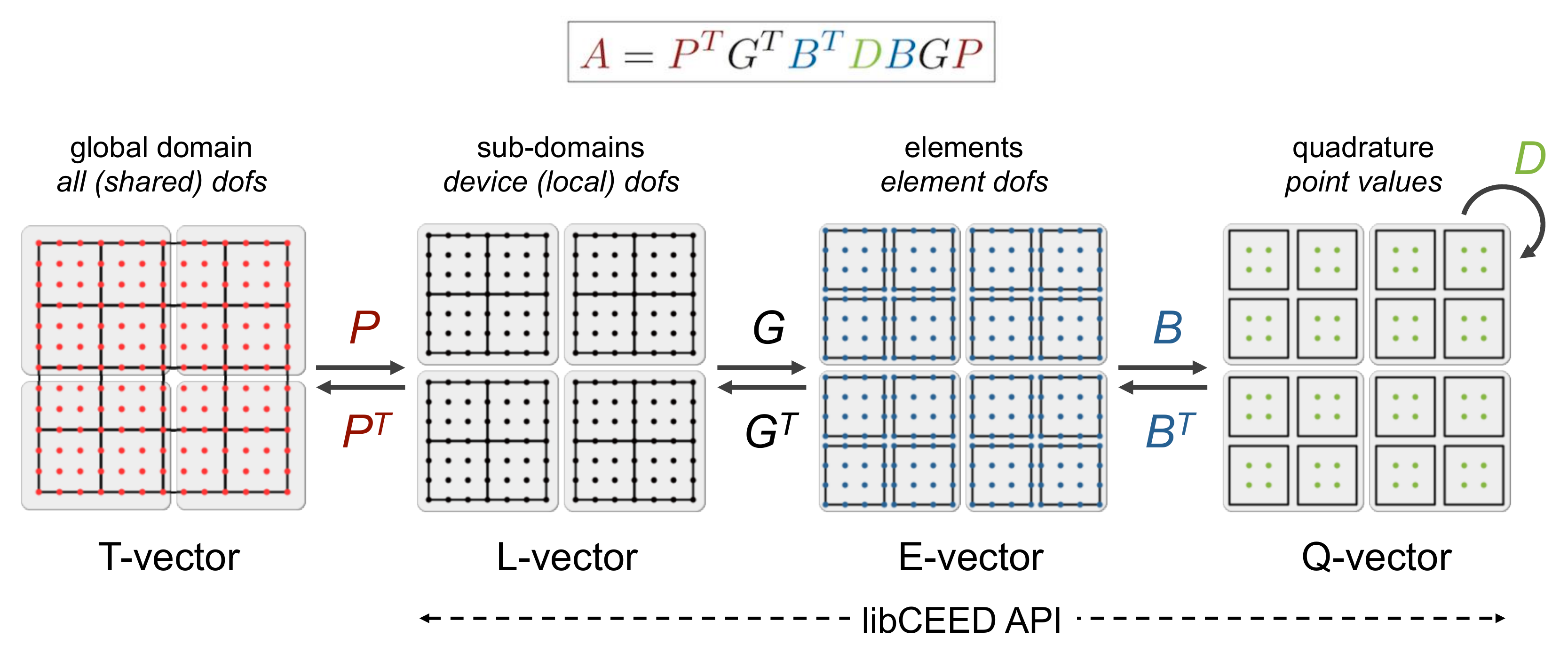}
\caption{\label{fig:operator-decomposition} Finite element operator
  decomposition}
\end{figure*}

This is illustrated below for the simple case of symmetric linear operator on
third order ($Q_3$) scalar continuous ($H^1$) elements, where we use the notions
\textbf{T-vector}, \textbf{L-vector}, \textbf{E-vector}, and \textbf{Q-vector}
to represent the sets corresponding to the (true) degrees of freedom on the
global mesh, the split local degrees of freedom on the subdomains, the split
degrees of freedom on the mesh elements, and the values at quadrature points,
respectively (see Figure \ref{fig:operator-decomposition}).
We refer to the operators that connect the different types of vectors as:
\begin{itemize}
\item $\boldsymbol{P}$: Subdomain restriction
\item $\boldsymbol{G}$: Element restriction
\item $\boldsymbol{B}$: Basis (DOFs-to-Qpts) evaluator
\item $\boldsymbol{D}$: Operator at quadrature points
\end{itemize}

More generally, when the test and trial space differ, they have their own
versions of $\boldsymbol{P}$, $\boldsymbol{G}$ and $\boldsymbol{B}$.

The libCEED API takes an algebraic approach, where the user describes in the
\textit{frontend} the operators $\boldsymbol{G}$, $\boldsymbol{B}$, and
$\boldsymbol{D}$ and the library provides \textit{backend} implementations and
coordinates their action to the original operator on \textbf{L-vector} level
(i.e. independently on each device / MPI task). The subdomain restriction
operation, $\boldsymbol{P}$ is outside of the scope of the current libCEED API.

One of the advantages of this purely algebraic description is that it includes
all the finite element information, so the backends can operate on linear
algebra level without explicit finite element code. The frontend description is
general enough to support a wide variety of finite element algorithms, as well
as some other types algorithms such as spectral finite differences. The
separation of the front and backends enables applications to easily switch/try
different backends and enables backend developers to impact many applications
from a single implementation.

The mapping between the decomposition concepts and the code implementation is as
follows:

\begin{itemize}
\item \textbf{L-}, \textbf{E-}, and \textbf{Q-vectors} are represented by
  \textit{CeedVector} objects\footnote{A backend may choose to operate
    incrementally without forming explicit \textbf{E-} or \textbf{Q-vectors}.}
\item $\boldsymbol{G}$ is represented as a \textit{CeedElemRestriction}
\item $\boldsymbol{B}$ is represented as a \textit{CeedBasis}
\item $\boldsymbol{D}$ is represented as a \textit{CeedQFunction}
\item $\boldsymbol{G}^T \boldsymbol{B}^T \boldsymbol{D} \boldsymbol{B}
  \boldsymbol{G}$ the local action of the operator is represented as a
  \textit{CeedOperator}
\end{itemize}

Users can provide source code for pointwise application of their weak form in a
single source file using mutually supported constructs from C99, C++11, and
CUDA.

\begin{figure}[ht!]
\centering
\includegraphics[width=\columnwidth]{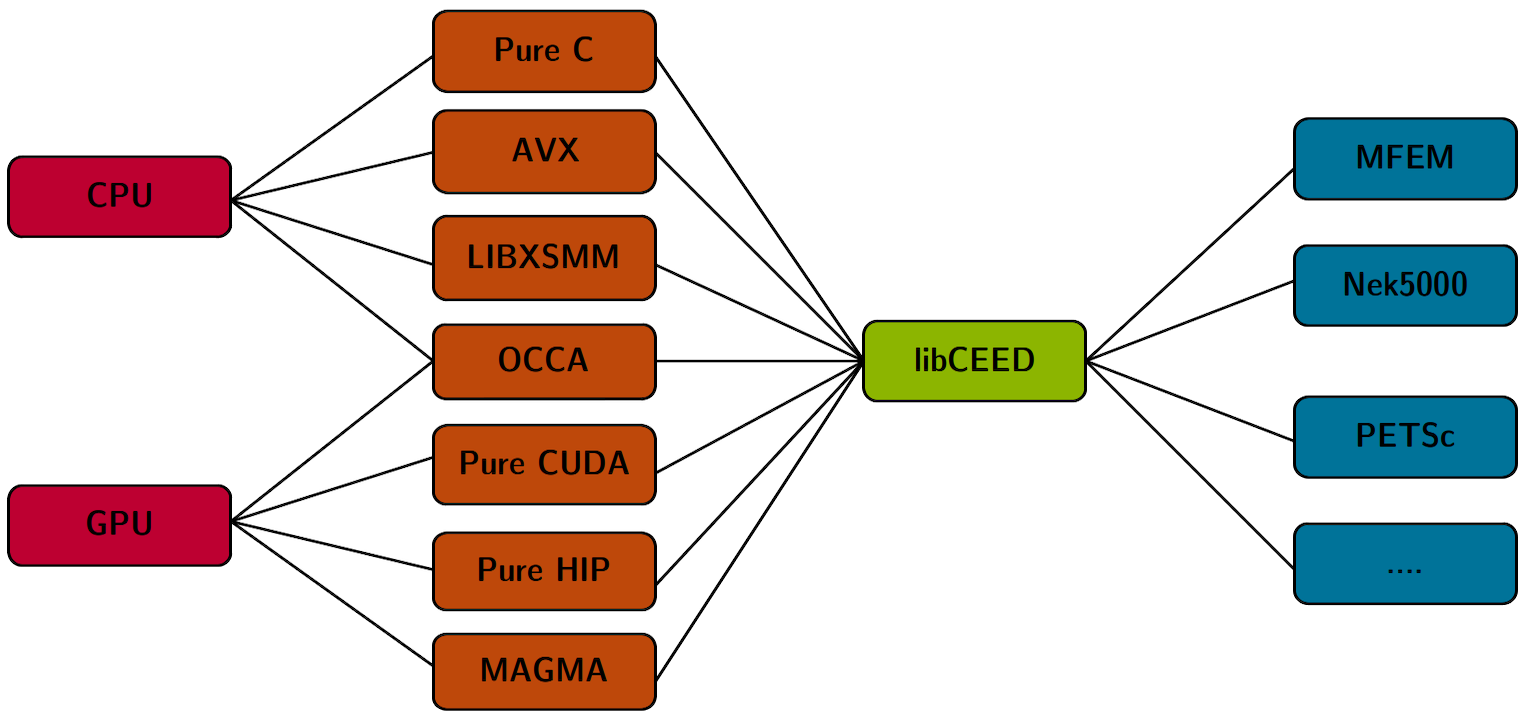}
   \caption{\label{fig:api-backends}
            The role of libCEED as a low-level API.}
\end{figure}

\subsection*{CPU and GPU Performance}

libCEED provides a unified interface for all types of hardware, allowing users
to write a single source code and to select the desired backend at run time.
Backends differ in the hardware they target, but also in their implementation
and algorithmic choices.

libCEED provides backends for CPUs, NVIDIA GPUs, and AMD GPUs implemented in C
(with or without AVX intrinsics), CUDA, and HIP respectively. libCEED also
provides backends taking advantage of specialized libraries like libXSMM for
CPUs, or MAGMA for NVIDIA and AMD GPUs (see Figure \ref{fig:api-backends}). The
OCCA backend is special in the sense that it aims at supporting all possible
hardware in a unified backend.

Backends are interoperable, allowing to use different backends together on
heterogeneous architectures. Moreover, each process or thread can instantiate an
arbitrary number of backends, this can be used to select the backend with the
highest performance for each operator, or to run on mixed meshes.

The best performing CPU backends use the LIBXSMM library. In order to best use
modern CPU architectures, the basis application operations are decomposed as
small matrix multiplications, using tensor contractions when on tensor product
bases. LIBXSMM provides efficient computation of these small matrix-matrix
products, vectorized across the quadrature points for a single element or
batches of elements.

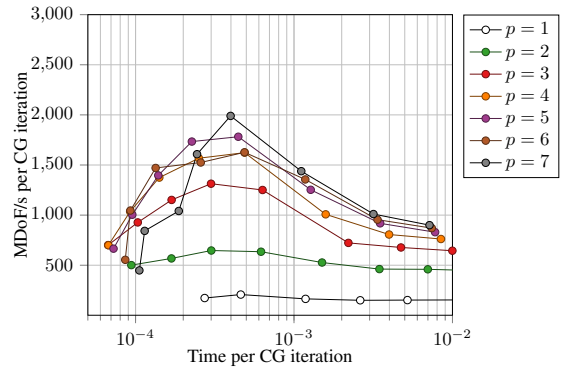
\begin{figure}[ht!]
\begin{tikzpicture}[scale=0.7]
  \begin{axis}[
  xmode=log,
  grid=both,
  major grid style={line width=.1pt,draw=gray!50},
  minor grid style={line width=.1pt,draw=gray!50},
  cycle list name=will,
  log basis x={10},
  tick align=outside,
  tick pos=left,
  xmin=5e-05, xmax=0.01,
  ymin=1,
  ymax=3000,
  xlabel={Time per CG iteration},
  ylabel={MDoF/s per CG iteration},
  legend cell align={left},
  legend pos = outer north east,
  legend entries = {$p=1$,$p=2$,$p=3$,$p=4$,$p=5$,$p=6$,$p=7$},
  ]
  \addplot
table {%
0.000272989 172.721
0.00046134 207.433
0.00118279 164.425
0.00261807 150.18
0.00520134 152.287
0.0103836 153.778
0.0211196 152.024
0.0443032 145.466
};
\addplot
table {%
9.4085875e-05 501.148
0.000168508888888889 567.905
0.00030067625 646.813
0.0006193225 634.86
0.00150291111111111 527.043
0.00345995555555556 461.5
0.00699034 459.304
0.01445074 445.971
};
\addplot
table {%
6.71745e-05 701.1
0.000103271 927.114
0.000168884 1151.56
0.0002995135 1313.4
0.000633075 1250.69
0.002204 722.845
0.004733145 677.199
0.00998674 644.74
};
\addplot
table {%
6.7413e-05 699.436
9.24585e-05 1036.39
0.000141227 1374.4
0.0002507925 1567.76
0.000487983 1623.21
0.001584565 1007.7
0.00398134 806.434
0.00846114 761.672
};
\addplot
table {%
7.26105e-05 665.387
9.53795e-05 1003.82
0.000139141 1396.45
0.0002266765 1733.85
0.0004450085 1781.95
0.00127474 1252.56
0.003497075 918.011
0.00777292 829.658
};
\addplot
table {%
8.6105e-05 554.034
9.28045e-05 1045.8
0.000133896 1470.42
0.000257945 1524.95
0.0004871845 1626.79
0.001178085 1355.83
0.003366075 954.379
0.00740395 870.245
};
\addplot
table {%
0.000105679 448.49
0.000114131 842.769
0.0001874685 1041.25
0.000244355 1607.63
0.00039804 1989.75
0.001111255 1438.37
0.003172255 1010.69
0.00716405 899.917
};
  \end{axis}
  \end{tikzpicture}
  \caption{\label{fig:perfsCPU}Throughput vs latency for the libCEED
    \code{/cpu/self/xsmm/blocked} backend solving BP3 on a 2-socket AMD EPYC
    7452.}
\end{figure}

The best performing GPU backends on the CEED benchmark problems are
\code{/gpu/cuda/gen} for quadrilateral and hexahedral elements, and
\code{/gpu/magma} for simplex elements.

The \code{/gpu/cuda/gen} backend is using runtime code generation and JIT
compilation to generate a unique optimized GPU kernel for each libCEED operator.
The \code{gpu/magma} backend is based on the MAGMA library.

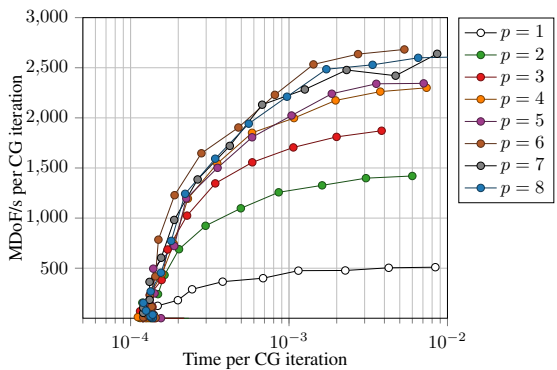
\begin{figure}[ht!]
\begin{tikzpicture}[scale=0.7]
  \begin{axis}[
  xmode=log,
  grid=both,
  major grid style={line width=.1pt,draw=gray!50},
  minor grid style={line width=.1pt,draw=gray!50},
  cycle list name=will,
  log basis x={10},
  tick align=outside,
  tick pos=left,
  xmin=5e-05, xmax=0.01,
  ymin=1,
  ymax=3000,
  xlabel={Time per CG iteration},
  ylabel={MDoF/s per CG iteration},
  legend cell align={left},
  legend pos = outer north east,
  legend entries = {$p=1$,$p=2$,$p=3$,$p=4$,$p=5$,$p=6$,$p=7$,$p=8$},
  ]
  \addplot
  table {%
  0.000218933 0.123325
  0.000169932 0.264812
  0.000160804 0.466405
  0.000152783 0.818155
  0.000119983 1.87527
  0.000114926 3.52399
  0.000125232 5.8212
  0.000118245 11.6453
  0.00012832 20.2696
  0.000140066 35.0763
  0.000134106 71.1155
  0.000148314 124.823
  0.000198899 180.679
  0.000244523 289.482
  0.000381226 365.728
  0.000686071 400.287
  0.00114595 475.61
  0.00226555 477.441
  0.00425817 504.134
  0.00838328 510.151
  };
  \addplot
  table {%
  0.000233264 0.115749
  0.00017693 0.254339
  0.000161957 0.463086
  0.000150167 0.832409
  0.000134685 1.67056
  0.000128243 3.15807
  0.000114343 6.37557
  0.000127604 10.7912
  0.000115828 22.4557
  0.000121849 40.3204
  0.000126797 75.215
  0.000118709 155.953
  0.000148962 241.249
  0.000163785 432.183
  0.000202206 689.521
  0.000297476 923.184
  0.000496519 1097.69
  0.000860303 1257.31
  0.00161803 1326.73
  0.00305839 1398.36
  0.00599976 1420.11
  };
  \addplot
  table {%
  0.000217329 0.294484
  0.000150536 0.74401
  0.000143479 1.36606
  0.000119664 2.86636
  0.000118739 5.36472
  0.000129313 9.14838
  0.000115348 19.0467
  0.000122816 34.4011
  0.000114816 70.7655
  0.000137425 113.698
  0.000132138 231.765
  0.000157046 382.212
  0.000171259 686.966
  0.000227257 1024.82
  0.000342473 1346.21
  0.000586435 1556.31
  0.00106503 1705.06
  0.00199606 1810.14
  0.00384129 1871.52
  };
  \addplot
  table {%
  0.000142183 0.879152
  0.000125111 1.7984
  0.000131197 3.08697
  0.000115801 6.29526
  0.000111198 12.3833
  0.00012863 20.2207
  0.000136184 36.0761
  0.000124486 76.611
  0.000120667 153.422
  0.000135837 264.56
  0.000158324 447.091
  0.000187077 745.282
  0.000230045 1193.79
  0.000352755 1545.05
  0.000584391 1850.93
  0.00107462 1997.63
  0.00196724 2173.98
  0.00376693 2261.88
  0.0073803 2299.99
  };
  \addplot
  table {%
  0.000155554 1.38858
  0.000140224 2.82406
  0.000138402 5.2456
  0.000121662 10.9401
  0.00012309 20.6434
  0.000121672 39.8695
  0.000135322 68.4366
  0.000134195 134.737
  0.000141732 249.069
  0.000139071 495.58
  0.000188863 720.951
  0.000224859 1196.31
  0.000353953 1501.45
  0.000584862 1806.1
  0.00103758 2023.56
  0.00186235 2240.86
  0.00355595 2339.92
  0.00707707 2344.13
  };
  \addplot
  table {%
  0.000138978 2.46801
  0.00014393 4.42575
  0.000120322 9.83192
  0.000135392 16.2269
  0.00012845 32.8922
  0.000135226 60.0847
  0.0001359 114.974
  0.000132513 231.11
  0.000143561 418.114
  0.00014988 784.957
  0.000189588 1228.44
  0.000279983 1646.68
  0.00047939 1903.82
  0.000814448 2229.65
  0.00142627 2533.28
  0.00272713 2636.13
  0.00534601 2682.53
  };
  \addplot
  table {%
  0.000137165 3.73273
  0.000136171 7.04996
  0.000135109 13.3226
  0.000136813 24.6688
  0.000120211 54.2794
  0.000120276 104.883
  0.00013198 184.793
  0.00013178 363.765
  0.000155967 604.109
  0.000188632 981.768
  0.000264976 1385.55
  0.000422989 1720.69
  0.000676972 2131.4
  0.00125825 2283.35
  0.002309 2477.53
  0.00470555 2420.68
  0.00861265 2639.21
  };
  \addplot
  table {%
  0.000138315 5.27058
  0.000134813 10.2141
  0.000133527 19.4791
  0.000137809 35.6507
  0.000125164 76.1961
  0.000120419 153.738
  0.000133981 268.224
  0.000155618 454.863
  0.000180491 772.477
  0.000221054 1242.34
  0.00034221 1592.66
  0.000556995 1941.97
  0.000970898 2211.03
  0.00172093 2485.13
  0.00336881 2529.18
  0.00653128 2598.97
  0.0129847 2609.47
  };
  \end{axis}

  \end{tikzpicture}
  \caption{\label{fig:perfsGPU}Throughput vs latency for the libCEED
    \code{/gpu/cuda/gen} backend solving BP3 on a NVIDIA V100.}
\end{figure}

In order to explore the weak and strong scaling of these algorithms on CPU and
GPU architectures, we consider the throughput in terms of the latency. High
throughput for low latencies are good for architectures and problems that need
strong scalability. On the other hand, high throughput for high latency
corresponds to architectures that weak-scale efficiently. Comparing these
measures for a 2-socket AMD EPYC 7452 CPU, in Figure~\ref{fig:perfsCPU}, with an
NVIDIA V100 GPU in Figure~\ref{fig:perfsGPU}, using the best CPU and GPU backends
of libCEED, we observe a much better ability for the CPU to strong scale over
the GPU, but a better weak scalability for the NVIDIA V100 over the AMD EPYC
7452 on the benchmark problem BP3 (in both Figures, $p$ is the polynomial order).

For all GPU backends the weak and strong scalability behave similarly.
A constant GPU overhead limits dramatically the strong scalability.
Any kernel below a million degrees of freedom is dominated by this constant cost,
therefore the computation time between one and a million degrees of freedom is
roughly the same, which correspond to the clustering on the left of
Figure~\ref{fig:perfsGPU}.
However, above a million degrees of freedom, we can observe a superlinear weak
scalability: a higher number of degrees of freedom results in higher
performance per degree of freedom.
The CPU backends have typical CPU performance profile, with decent strong and
weak scalability, and optimal performance achieved in the middle for a number of
degrees of freedom that depends on the architecture cache memories, see
Figure~\ref{fig:perfsCPU}.

%% file: tex/nek+mfem.tex
\section{Nek and MFEM} \label{sec:NekMFEM}

At a higher level of abstraction, CEED provides a ``high-level API'' to
applications through the MFEM and Nek discretization libraries. This API
operates with global discretization concepts, specifying a global mesh, finite
element spaces and PDE operators to be discretized with the point-wise physics
representing the coefficients in these operators. Given such inputs, CEED
provides efficient discretization and evaluation of the requested operators,
without the need for the application to be concerned with element-level
operations. Internally, the high-level API can make use of CEED's low-level APIs
described in the previous sections. The global perspective also allows CEED
packages to provide general unstructured adaptive mesh refinement support, with
minimal impact in the application code.

\subsection*{Nek5000/CEM/RS}

Nek5000 is a thermal-fluids code based on the spectral element method (SEM)
\citep{pat84} that is used for a wide range of scientific applications,
including reactor thermal-hydraulics, thermal convection, ocean modeling,
combustion, vascular flows, and fundamental studies of turbulence. NekCEM
supports both an SEM and an SE discontinuous-Galerkin (SEDG) formulation for
applications in electromagnetics, drift-diffusion, and quantum-mechanical
systems. These codes have scaled to millions of MPI ranks using the Nek-based
{\em gslib} communication library to handle all near-neighbor and other stencil
type communications (e.g., for algebraic multigrid). Tensor contractions
constitute the principal computational kernel, which leads to high CPU
performance with a minor amount of tuning.

\begin{figure}[!ht]
\centering
\includegraphics[width=0.8\textwidth]{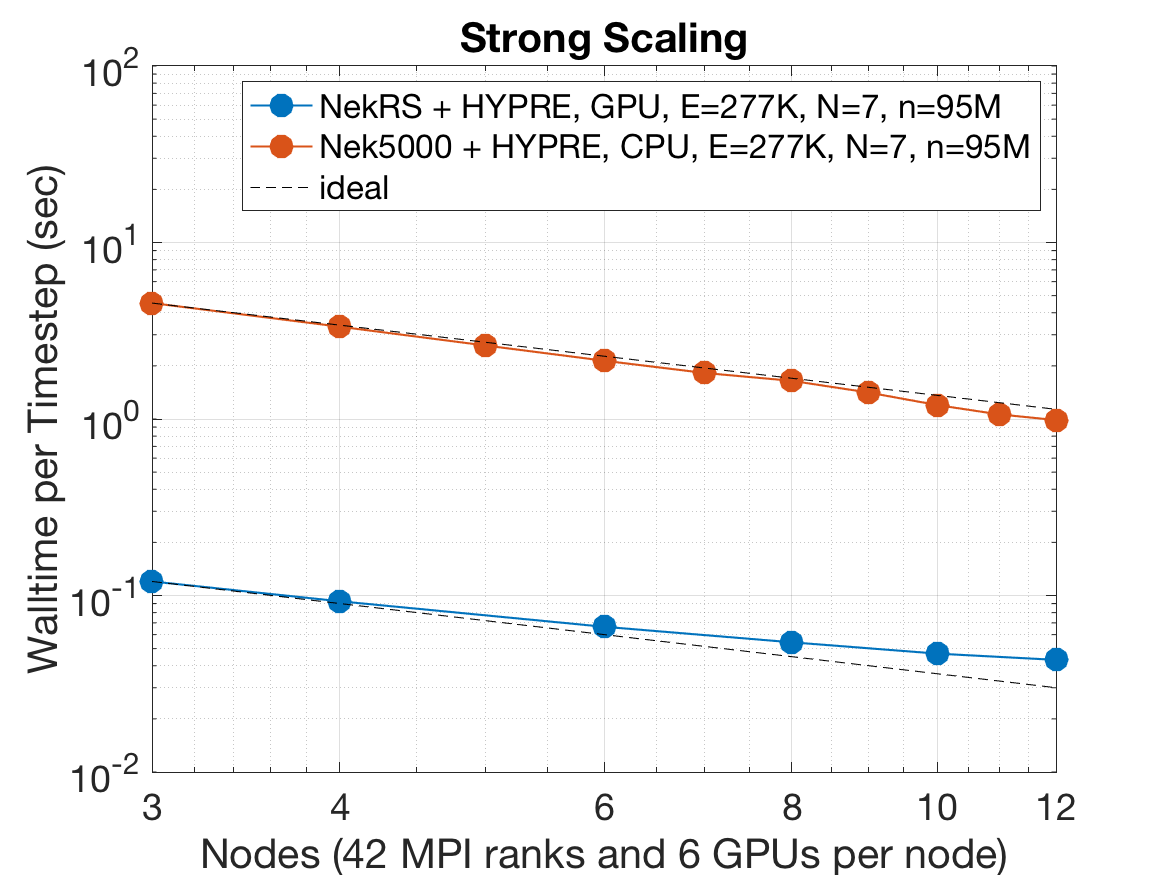}
\vskip.1in
\includegraphics[width=0.8\textwidth]{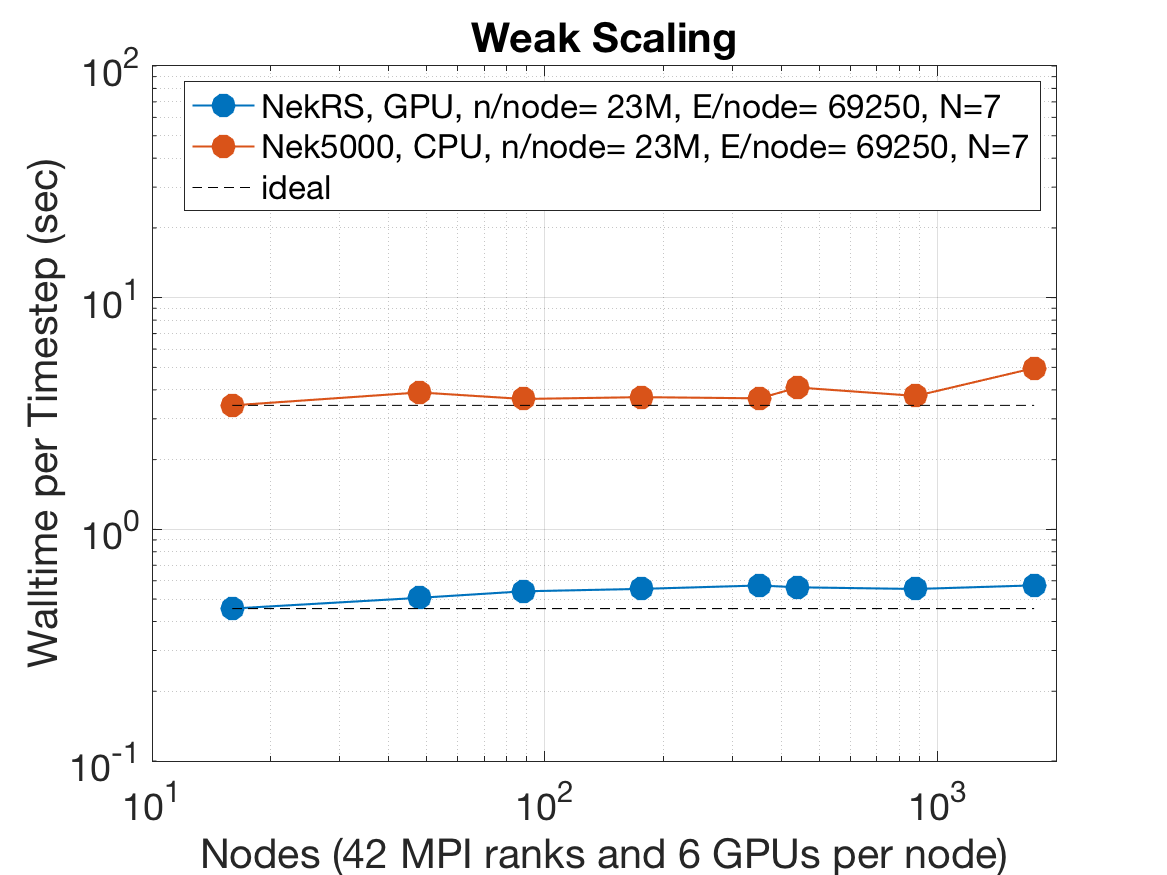}
\caption{NekRS GPU and Nek5000 CPU strong- and weak-scaling performance on
  OLCF's Summit. The plots show the wall-time per step, averaged over 100 steps,
  for turbulent flow in a 17$\times$17 rod-bundle example coming from ExaSMR
  (Figure 17a).}
\label{fig:nek-scaling}
\end{figure}

Initial GPU development and testing was done with NekCEM using OpenACC
\citep{min2015a}. For portability reasons, NekRS---the GPU variant of
Nek5000---was built on top of kernels from libParanumal using OCCA. In both
cases, node-level parallelism requires kernels written at a higher level than
simple tensor contractions. For performance, full operations (e.g., $\nabla u$)
are cast into a single kernel call for the GPU. Significant effort has gone into
overlapping the gather-scatter operation that is central to matrix-free SEM/FEM
operator evaluation. On a CPU platform, where there are only one or two spectral
elements per MPI rank, there is no opportunity for communication overlap.
However, GPUs such as the NVIDIA V100 require about 2 million gridpoints per
V100 for reasonable efficiency, which means that there is enough work on
subdomain interior points to cover some inter-node communication. Strong- and
weak-scaling performance for NekRS and Nek5000 on Summit are illustrated in
Figure \ref{fig:nek-scaling}. The strong-scaling plots reflect the most recent
performance enhancements in NekRS, including communication overlap and improved
preconditioners. Time-per-step in this case is less than 0.1 seconds.

A major push for Nek5000/CEM/RS applications is in the area of solvers.
Effective preconditioners for the Poisson problem are of primary importance for
the unsteady Navier-Stokes equations. Steady-state solvers are important for the
drift-diffusion equations and for Reynolds-averaged Navier-Stokes (RANS) models
used in nuclear engineering. Jacobi-free Newton-Krylov methods \citep{knoll04}
are under development for these applications.

\subsection*{The MFEM Finite Element Library}

MFEM is a free, lightweight, scalable C++ library for finite element methods
\citep{mfem,mfem-web}. Its goal is to enable high-performance scalable finite element
discretization research and application development on a wide variety of
platforms, ranging from laptops to exascale supercomputers. It also provides a
range of features beyond finite elements that allow for rapid prototyping and
development of scientific and engineering simulations. In CEED, MFEM is a main
component of the efforts in the Applications and Finite Element thrusts.

MFEM includes capabilities for basic linear algebra: vectors, dense and sparse
matrices and operations with them; iterative (Krylov) linear solvers; smoothers
and preconditioners, including multigrid; nonlinear operators and solvers; and
time stepping methods.
The library offers support for a wide variety of mesh types and operations on
them: arbitrary high-order curvilinear meshes in 1D, 2D (triangles and quads),
and 3D (tets, hexes, prisms), including surface and periodic meshes; mesh import
from meshing tools such as Gmsh, Netgen, CUBIT; adaptive conforming mesh
refinement for simplicial meshes; adaptive nonconforming refinement and
derefinement for all mesh types, including parallel rebalancing; mesh
optimization via node movement: TMOP \citep{dobrev2019target}.
The PDE discretization features include: arbitrary order $L^2$- (discontinuous),
$H^1$- (continuous), $H(\mathrm{div})$-, and $H(\mathrm{curl})$-conforming
finite elements and discretization spaces; NURBS meshes and discretization
spaces (IGA); a large variety of predefined linear, bilinear, and nonlinear
forms; support for many discretization approaches including continuous, mixed,
DG, DPG, IGA, etc.
In terms of parallel programming, MFEM supports MPI-based distributed memory
parallelism, OpenMP-based shared memory parallelism on CPUs, and
GPU-acceleration through various backends (see below).
Last but not least, the source distribution includes many examples and miniapps
that can be used as an introduction to the library and its capabilities, as well
as templates for developing more complex simulations.

In addition to its built-in capabilities, MFEM provides integration with many
other scientific libraries, including ECP software technologies projects such as
\emph{hypre}, PETSc, SUNDIALS, PUMI, libCEED, OCCA, etc. Support for the GPU
capabilities in some of these libraries is already available (e.g.\ OCCA,
libCEED) and for others it is currently under active development (e.g.\ \emph{hypre},
SUNDIALS).

Starting with version 4.0 (released in May 2019), MFEM introduced initial
support for GPU accelerators. Since then these capabilities are being actively
developed to add support in more components of the library while also improving
the performance of already existing kernels. The set of examples and miniapps in
MFEM that support GPUs is growing and now includes a number of PDE problems:
diffusion, advection, definite Maxwell, grad--div, Darcy, etc. Other algorithms
like AMR, TMOP \citep{dobrev2019target}, and multigrid are also supported (at
least partially) on GPUs.

\begin{figure*}[!ht]
\centering
\includegraphics[width=0.8\textwidth]{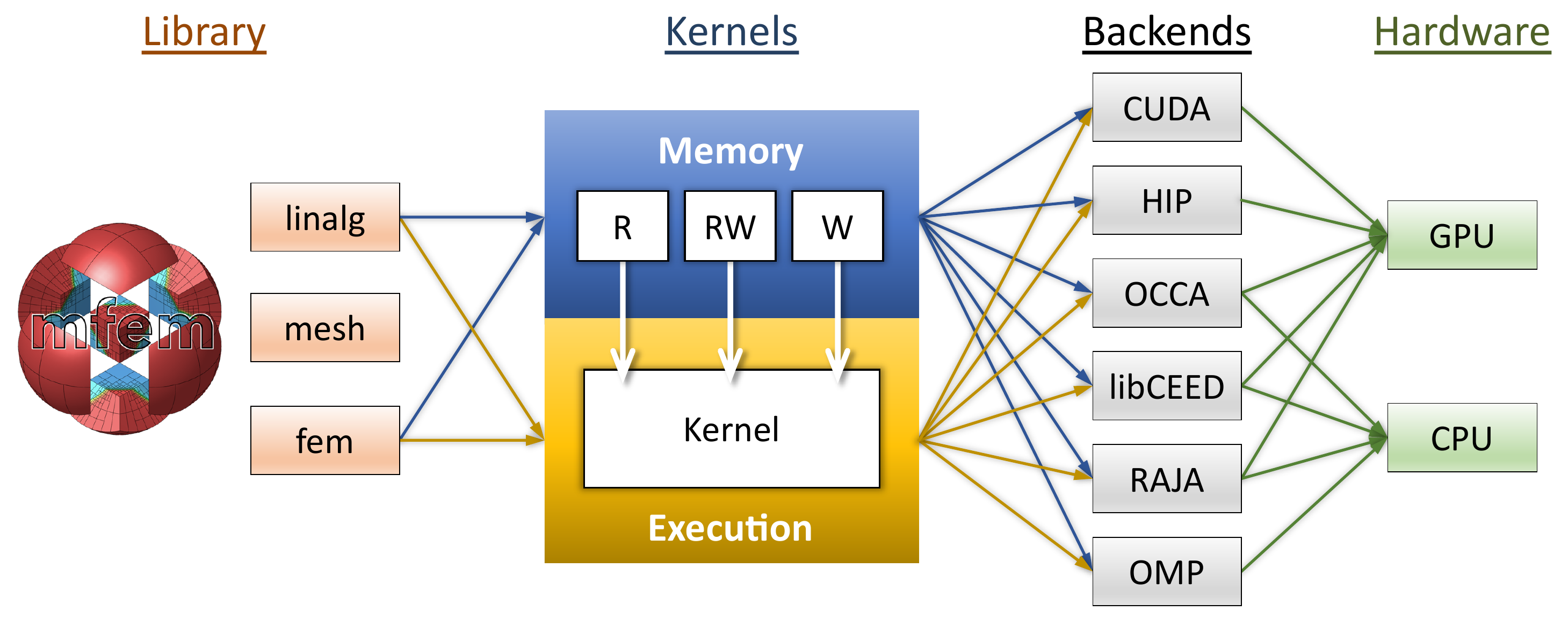}
\caption{Conceptual diagram of MFEM's portability abstractions.
\label{fig:mfem-diagram}}
\end{figure*}

The support for different hardware (CPUs and GPUs) and different programming
models (such as CUDA, OpenMP, HIP, RAJA \citep{RAJA-paper,RAJA-web}, OCCA,
libCEED) is facilitated by the concept of \emph{backends}, see Figure
\ref{fig:mfem-diagram}. The selection of the backend happens at runtime at the
start of the program which allows code to be developed, tested, and used without
the need to recompile the library or the application. The backends currently
supported (specified as strings) are: \code{cpu}, \code{raja-cpu},
\code{occa-cpu}, \code{ceed-cpu}, \code{omp}, \code{raja-omp}, \code{occa-omp},
\code{debug}, \code{hip}, \code{cuda}, \code{raja-cuda}, \code{occa-cuda}, and
\code{ceed-cuda}.

To facilitate gradual transition to GPU architectures for users and for the
library itself, MFEM introduced two features: a lightweight memory manager to
simplify the handling of separate host and device memory spaces, and a set of
\code{MFEM\_FORALL} macros for writing portable kernels that can dispatch
execution to different backends.

\begin{figure}[!ht]
\centering
\includegraphics[width=0.49\textwidth]{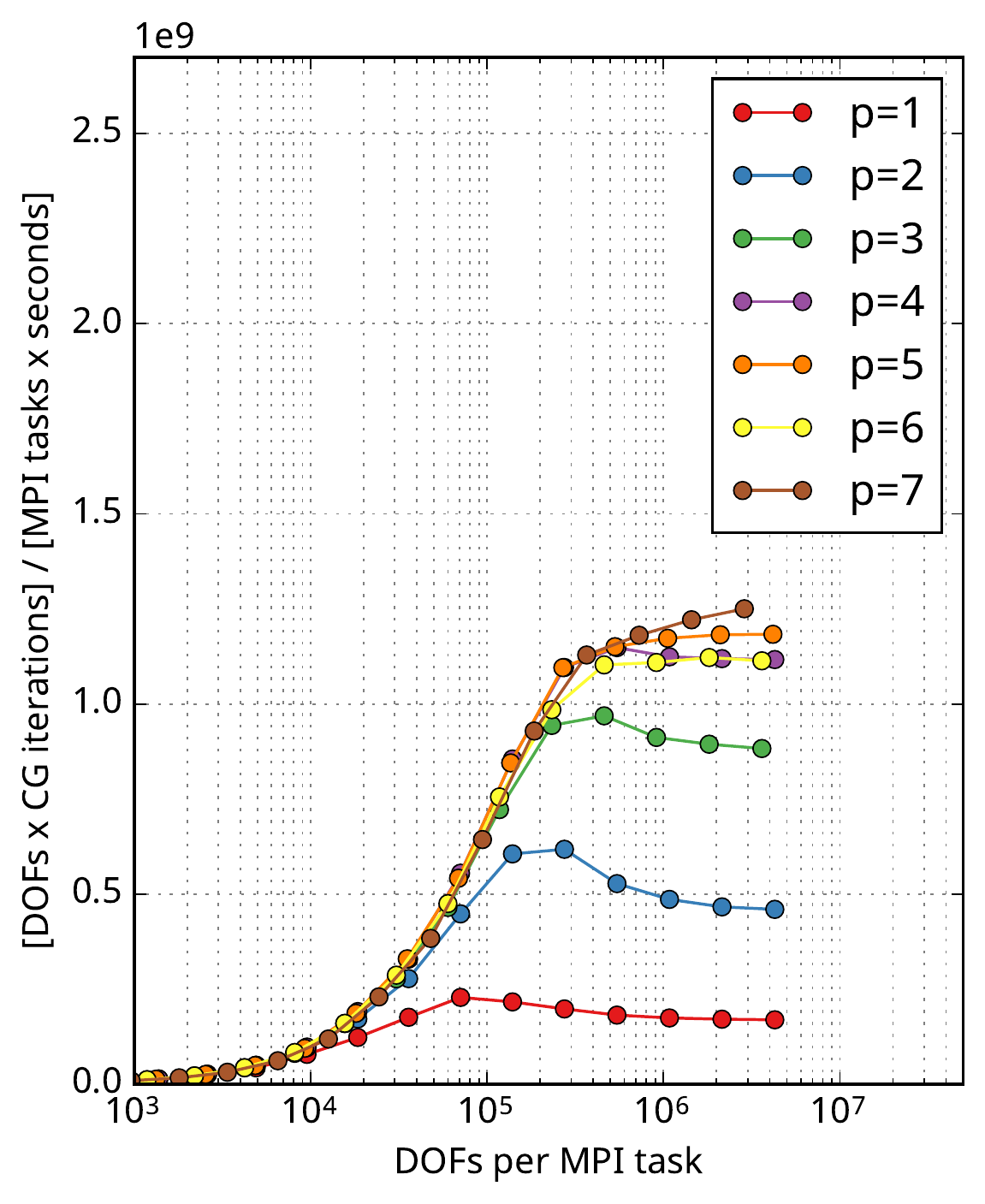}
\includegraphics[width=0.49\textwidth]{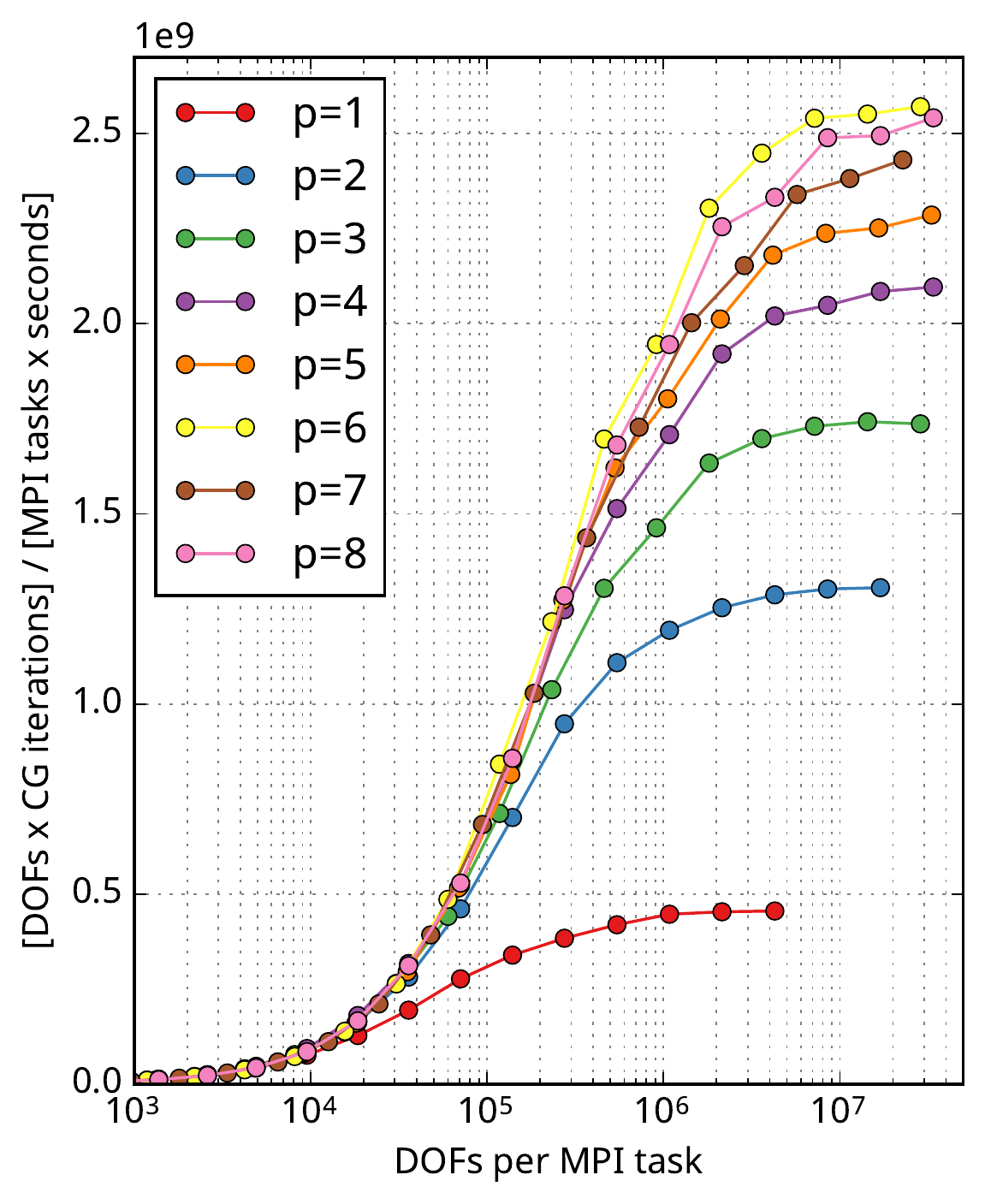}
\caption{BP3 performance comparison of MFEM's \code{cuda} (left) and \code{ceed-cuda}
(i.e. \code{/gpu/cuda/gen} from libCEED, right) backends on a single V100 GPU (Lassen
machine at LLNL).
\label{fig:mfem-gpu-results}}
\end{figure}

These capabilities are illustrated in Figure \ref{fig:mfem-gpu-results}, where
we present results for the BP3 benchmark (implemented as a slightly modified
version of MFEM's Example 1) on a single V100 GPU on LLNL's Lassen
machine. These results show the performance advantage of libCEED's CUDA-gen
backend (exposed in MFEM as the \code{ceed-cuda} backend) over the \code{cuda}
MFEM backend. The main reason for this improvement is the additional kernel
fusion used by CUDA-gen: the action of the operators $G^T B^T D B G$ (see
Section \ref{sec:libceed}, and Figure \ref{fig:operator-decomposition}) is
implemented as one kernel whereas the \code{cuda} backend uses three separate
kernels for $G^T$, $B^T D B$, and $G$. Another MFEM result using the
\code{ceed-cuda} backend to solve BP3 on 1024 V100 GPUs on Lassen was presented
earlier in the right panel of Figure \ref{fig:bps_ecp}.

%% file: tex/applications.tex
\section{Application Integration} \label{sec:applications}

\begin{figure*}[ht]
\begin{center}
\includegraphics[width=1.0\textwidth]{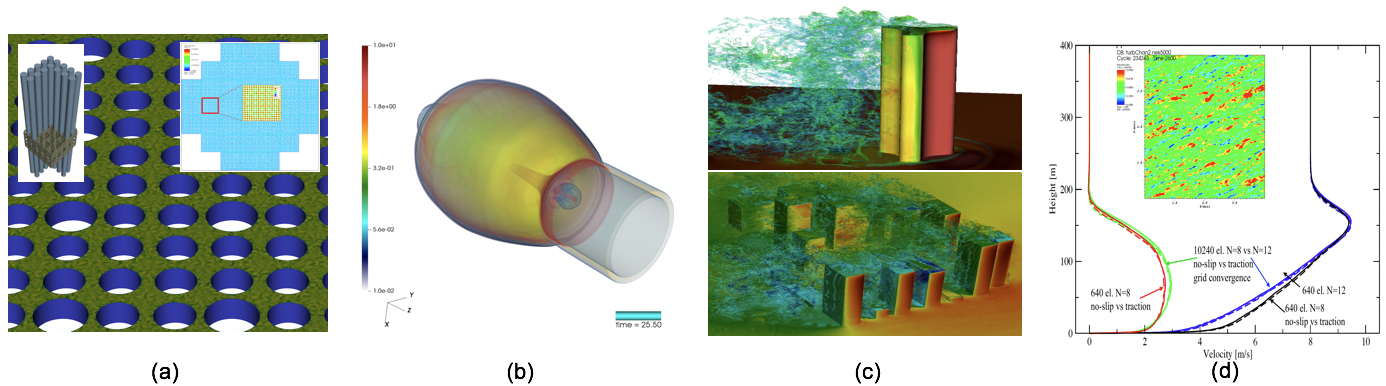}
\caption{ECP applications:
 (a) ExaSMR: $17\times 17$ rod-bundle turbulent flow simulation.
 (b) MARBL: 3D multi-material ALE simulation that is used as a performance benchmark.
 (c) Urban: LES modeling for vortex flows around Lake Point Tower and 20 buildings
     in Chicago downtown block.
 (d) ExaWind: GABLS benchmark studies with no-slip and traction boundary conditions.}
\label{ecp-apps}
\end{center}
\end{figure*}

The ultimate goal of CEED is to extend state-of-the-art high-order algorithms to
DOE ECP mission applications. This section illustrates the use of CEED-developed
technologies for several ECP applications including ExaSMR, MARBL, Urban, and
ExaWind. We also demonstrate impact over a range of other important applications
including work sponsored by DOE's Nuclear Energy Advanced Modeling and
Simulation program, Vehicle Technologies Office, COVID-19 research, and
SciDAC. Applications in these areas present significant challenges with respect
to scale resolution, multiphysics, and complex computational domains. The CEED
team has focused on developing algorithmic and scientific research at scale to
address these issues in collaboration with the application teams. The outcomes
of the CEED technologies have been integrated into the open source codes,
Nek5000/CEM/RS, MFEM, libCEED and libParanumal. Their impact in various
application problems have been demonstrated in the CEED milestone reports
\citep{ceed-ms1,ceed-ms8,ceed-ms20,ceed-ms23,ceed-ms29,ceed-ms32,ceed-ms34}
and other project reports \citep{nek-exasmr2017,anl-neams2019,nek-vto2020}.
Some of the results are shown in Figures~\ref{ecp-apps}--\ref{nonecp-apps}.

\subsection*{Small Modular Reactor Analysis: ExaSMR}
The goal of the ExaSMR project is to combine advanced thermal-hydraulics
modeling with scalable neutronics computations to allow reactor-scale modeling.
For the thermal-hydraulics analysis, the reactor-core comprises hundreds of
thousands of channels supporting turbulent flow with very fine solution scales.
The channels are typically hundreds of hydraulic diameters in length. For full
reactor-core simulations, Reynolds-Averaged Navier Stokes (RANS) approach in the
majority of the core with more detailed large eddy simulations (LES) is required
in critical regions. In addition, while the turbulence is challenging to
resolve, it tends to reach a statistically fully developed state within just a
few channel diameters, whereas thermal variations take place over the full core
size. This poses a challenge for coupled calculations. It is too expensive to
consider performing full large-eddy simulations (LES).

To accelerate the time-to-solution, CEED developed fully implicit and steady
state solvers for spectral-element-based thermal transport and RANS. For the
nonlinear Navier-Stokes and RANS transport, the Jacobian-free Newton Krylov
(JFNK) routines from NekCEM's drift-diffusion solver \citep{pht2020} have been
imported to Nek5000 and tested on various flow problems including vortex flow,
Dean's flow, lid-driven cavity, flow past cylinder and flow around rods
\citep{ceed-ms18}. This new steady-state solver uses an inexact (Jacobi-free)
formulation based on a first-order Taylor series expansion and converges to the
steady state solutions with a small number of pseudo-time steps.
Preconditioning the GMRES routine within the Newton step remains as future work.

While the thermal development time is governed by the long channel length, the
velocity rapidly reaches a statistically steady state, proportional to the
hydraulic diameter. By freezing the expensive-to-generate velocity field, one
can accelerate equilibration of the thermal field without having to laboriously
compute tens of thousands of transient turbulent eddies. We have developed
preconditioning strategies for steady or implicit advection-diffusion and
Navier-Stokes equations using tensor-product-based spectral element methods.
For the advection-diffusion problem, $p$-multigrid (PMG) is used directly as a
preconditioner within a Krylov subspace projection (KSP) method such as GMRES
\citep{ceed-ms20,pablo2019,Pazner2018e}. For Navier-Stokes, PMG is part of a
larger preconditioner that includes restriction of velocity search directions to
the space of divergence-free fields through a projection technique
\citep{pablo2020}. These strategies have been applied for turbulent
thermal-stress models for rod bundle simulations \citep{javier2019}.

For the unsteady ExaSMR simulations, the GPU-based NekRS code has made
significant advances in development. For the $17\times 17$ rod-bundle in
Figure~\ref{ecp-apps}(a), we have demonstrated improved NekRS simulation
capabilities to extend to largest problem size to date, with 175 million
spectral elements ($n=61$ billion grid points) using 3520 nodes (21120 V100s) on
Summit as well as strong and weak scaling performance at large scale in
\citep{ceed-ms32,ceed-ms34}. (See Figure 14.)

\subsection*{Compressible Shock Hydrodynamics}
MARBL is a next-gen multi-physics simulation code being developed at LLNL. The
code targets multi-material hydrodynamics and radiation/magnetic diffusion
simulations, with applications in inertial confined fusion, pulsed power
experiments, and equation of state/material strength analysis. The goal of this
application is to enhance LLNL's modular physics simulation capabilities, with
increased performance portability and flexibility. One of the central features
of MARBL is an ALE formulation based on the MFEM-driven BLAST code
\citep{Anderson2018}, which solves the conservation laws of mass, momentum, and
energy. The BLAST code utilizes high-order finite element discretizations of
several physical processes on a high-order (curved) moving mesh. The method
consists of (i) a Lagrangian phase, where the multi-material compressible Euler
equations are solved on a moving mesh \citep{Dobrev2012, Dobrev2016}, (ii) a
remesh phase, which improves the mesh quality \citep{Dobrev2020}, and a field
remap phase that performs a conservative and monotone advection between two
meshes \citep{Anderson2015}.

The first major step towards improved efficiency in MARBL was the introduction
of matrix-free / partial-assembly based methods. The CEED-developed Laghos
miniapp played a critical role for that, as it exposed the main computational
kernels of BLAST's Lagrangian phase, without the additional overhead of
physics-specific code. Laghos introduced partial assembly versions for many of
BLAST's specific kernels, which were later directly used by the application. For
its more standard finite element operations, BLAST utilized MFEM's tensor-based
routines. These included partially assembled bilinear forms for mass, diffusion
and advection; tensor-based evaluation of finite element functions and their
gradients; matrix-free diagonal preconditioning; and other algorithms as
well. These methods were used extensively throughout the application's
Lagrangian and remap phases. Furthermore, the CEED team derived a matrix-free
version of MFEM's mesh optimization miniapp, which could also be used directly
by the remesh phase of the application.

The GPU port of MARBL/BLAST is exclusively based on the partial assembly
technology from CEED and the GPU support via the MFEM version 4.0 release. The
CEED team developed GPU versions of Laghos and the MFEM's mesh optimization
miniapps. GPU kernels from these miniapps, together with general MFEM finite
element operations as the ones mentioned above, could be used directly by the
MARBL code. Application-specific operations, on the other hand, are implemented
in MARBL, making use of the RAJA kernel abstractions and MFEM memory management,
GPU-friendly data structures, small dense matrix kernels, use of shared memory,
etc.

The current state of MARBL's GPU capability provides around $15\times$ speedup
on the main benchmark problem, which is a multi-material ALE simulation on a 3D
unstructured mesh, see Figure~\ref{ecp-apps}(b). This comparison uses 4 CPU
nodes (144 cores) of LLNL's \textit{rzgenie} machine versus 4 GPU nodes (16
GPUs) of LLNL's \textit{rzansel} machine. Broken over the ALE phases, the
observed speedups are $16\times$ in the Lagrange phase, $15\times$ in the remap
phase, and 6$\times$ in the mesh optimization phase.

\subsection*{Flow in Urban Environments}
The urban challenge problem considers the assessment of extreme heat events on
buildings in dense urban environments, with up to a few 1000 buildings being
modeled during an event. This challenge problem involves coupling of WRF (to
define initial weather conditions), Nek5000 (to model heat transfer near
buildings), and EnergyPlus (to model heat emissions and energy performance). In
collaboration with the ECP-Urban team, CEED team built spectral element meshes
and performed LES simulations of Lake Point Tower and Chicago downtown block
consisting of 20 buildings as shown in Figure~\ref{ecp-apps}(c)
\citep{ceed-ms23,ceed-ms29}. The 20-building mesh comprises $E = 143340$
spectral elements and its simulation with $N=13$ is performed using 1024 nodes
of ALCF/Mira (32768 MPI ranks). This effort has also generated interest from
other federal agencies outside of DOE.

\begin{figure*}[ht]
\begin{center}
\includegraphics[width=1.0\textwidth]{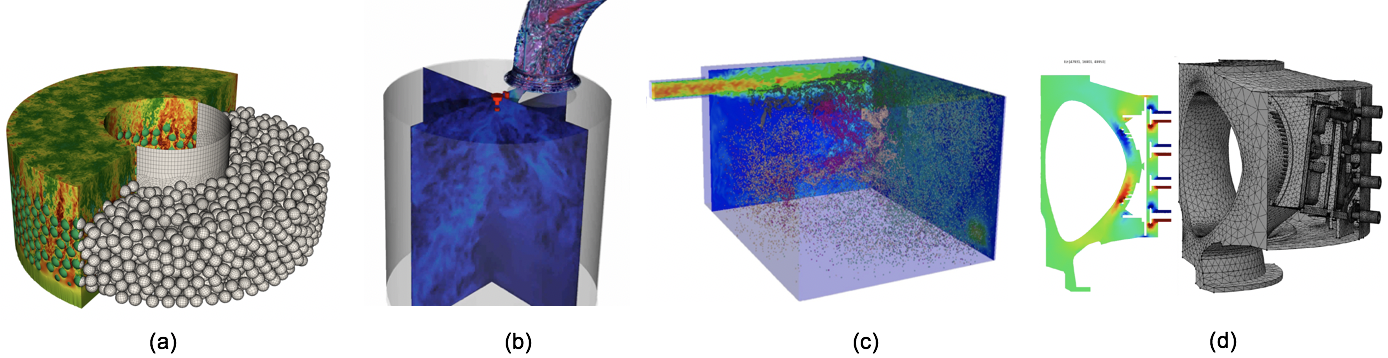}
\caption{Other applications:
 (a) DOE NEAMS: Turbulent flows around 3344 pebbles with an all-hex mesh.
 (b) DOE VTO: Exhaust stroke TCC engine modeling.
 (c) COVID19: LES Lagrangian particle tracking simulation for 500,000 aerosols.
 (d) SciDAC RF: EM analysis of the vacuum region from a RF antenna.}
\label{nonecp-apps}
\end{center}
\end{figure*}

\subsection*{Atmospheric Boundary Layer Flows}
Efficient simulation of atmospheric boundary layer flows (ABL) is important for
the study of wind farms, urban canyons, and basic weather modeling. In
collaboration with the ExaWind team, we identified an atmospheric boundary layer
benchmark problem~\citep{matt2013} to serve as a point of comparison for code
and modeling strategies. We have addressed cross-verification and validation of
our LES results and corresponding wall models. We demonstrated the suitability
of high-order methods for a well-documented stably stratified atmospheric
boundary layer benchmark problem, the Global Energy and Water Cycle Experiment
(GEWEX) Atmospheric Boundary Layer Study (GABLS) as shown in
Figure~\ref{ecp-apps}(d). This collaboration will be extended to perform scaling
studies to compare the performance of several ABL codes on CPU and GPU
platforms.

As another component of the ExaWind collaboration we performed RANS simulations
to compute the drag and lift forces of wind-turbine and NACA0012 airfoil
structures at Reynolds numbers up to $Re$=10 million
\citep{ceed-ms23,ceed-ms32}. We investigated several models for the boundary
layer treatment in the Nek5000 RANS solver including a wall-resolved regularized
approach where we have to use adequate resolution inside the very thin log and
viscous sub-layers \citep{ananias2018} and a stability-enhanced wall-resolved
$k$-$\omega$ and $k$-$\tau$ models where we do not need such high resolution.

\subsection*{Pebble-Bed Reactors}
Flow through beds of randomly-packed spheres is encountered in many science and
engineering applications. The meshing challenge is to have a high quality
all-hex mesh with relatively few elements per sphere. Working with the DOE NEAMS
project, the CEED team has developed novel scalable meshing strategies for
generating high-quality hexahedral element meshes that ensure accurate
representation of densely packed spheres for complex pebble-bed reactor
geometries. Our target is to capture highly turbulent flow structures in the
solution at minimal cost by using relatively few elements ($\approx$300 per
sphere) of high order ($p=7$). Algorithmic strategies including efficient edge
collapse, tessellation, smoothing, and projection, are presented in
\citep{ceed-ms34} along with quality measurements, flow simulations, validation,
and performance results for pebble bed geometries ranging from hundreds to
thousands of pebbles. Figure~\ref{ecp-apps}(a) shows a case of 3344 pebbles in
an annular domain using 1.1M spectral elements.

\subsection*{Internal Combustion}
Turbulence in IC engines presents a challenge for computational fluid dynamics
due primarily to the broad range of length and time scales that need to be
resolved. Specifically, simulations need to predict the evolution of a variety
of flow structures in the vicinity of complex domains that are moving. Executing
these simulations accurately and in a reasonable amount of time can ultimately
lead to engine design concepts with improved efficiency.

The CEED team has been working with researchers at ETH Zurich \citep{gg19} and
ANL's Energy Systems Division (under support from DOE's Vehicle Technologies
Office) on detailed studies of turbulence in the IC engine cycle. We developed a
characteristic-based spectral element method for moving-domain problems
\citep{patel18}, and demonstrated it for the TCC III engine model illustrated in
Figure~\ref{nonecp-apps}(b). We also added a significantly enhanced capability
for handling complex moving geometries by adding scalable support for overset
grids, referred to as NekNek, based on generalized Schwarz overlapping methods
\citep{mittal19b}. The NekNek multimesh coupling is based entirely on the
kernels in Nek's {\em gslib} communication library, which has scaled to millions
of MPI ranks. A newly developed preconditioner based on the SEM/FEM spectral
equivalence was shown to be effective for solving the pressure-Poisson systems
in these configurations \citep{Bello-Maldonado2019}.

\subsection*{Aerosol Transport Modeling}

Related to the current COVID-19 pandemic, the Nek5000 team, in collaboration
with NVIDIA and Utah State is researching aerosol transport analysis.
High-resolution LES coupled with Lagrangian particle tracking is used for
predicting the dynamics of virus-laden aerosols in indoor classroom
environments \citep{dutta20}. Figure~\ref{nonecp-apps}(c) demonstrates a recent
simulation, using 70 million grid points and 500,000 five-micron aerosols with
a future target of 1 billion polydisperse aerosols in a full classroom size.

This application uses efficient algorithms for point containment and general
interpolation in physical space, {\tt findpts} and {\tt findeval}, which are
available on CPU platforms in CEED's Nek5000 and MFEM codes. Detailed discussion
of these methods and their porting to exascale machines is beyond the scope of
this paper. Future developments may include synchronous utilization of the CPU
or particle tracking on the device.

\subsection*{Magnetic Fusion}

Accurate radio-frequency (RF) heating simulations of fusion systems like the
ITER tokamak require the definition of analysis domains that include detailed
antenna, reactor wall and physics region geometric representations. As is the
case with other wave equation simulations, the application of high-order
methods, with their higher rates of convergence and high flop rate to memory
access, is critical for the accurate simulation of these classes of problems.
The software components being integrated to address this simulation workflow as
part of the DOE SciDAC Center for Integrated Simulation of Fusion Relevant RF
Actuators \citep{RFSciDAC} must support higher order geometry and high order
analysis methods.

These components include complete curved domain definitions based on CAD system
produced models of RF antenna geometries and geometry construction tools for the
analysis domain that support defeaturing the antenna CAD models as desired, and
combining the CAD geometry with the reactor wall geometry and any other
``physics''. Historically, ad-hoc methods are employed to execute this time
consuming step. Recent efforts have focused on providing a graphical
construction tool for tokamak systems building on general geometry manipulation
capabilities \citep{simmetrix_web}. The user interface to the PetraM (Physics
EquationTranslator for MFEM) component \citep{shiraiwa2017rf} supports the
association of the RF simulation material properties, loads and boundary
conditions (essential and natural) to the analysis domain geometry. With these
in place, fully automatic mesh generation of an initial curved tetrahedral mesh
using either Gmsh \citep{gmsh_web} or MeshSim \citep{simmetrix_web} is
performed. If the curved meshes are not of sufficiently high order for the basis
functions to be used, a tool has been developed to increase the order of
approximation of mesh edges and faces on curved domain boundaries up to order
six. We then execute a high-order MFEM simulation supplemented with PetraM
routines to control needed field information to perform the RF analysis. This is
followed by error estimation and mesh adaptation using the conforming curved
mesh adaptation procedure in PUMI/MeshAdapt and return to the analysis step
until acceptable solution accuracy is obtained.

Figure~\ref{nonecp-apps}(d) shows an example of the application of the basic
steps in the workflow of a tokamak geometry with RF antenna geometry inserted.
The result shown is for a low order mesh. Recent results up to order five are
showing a clear advantage to use of higher order elements.

%% file: tex/conclusion.tex
\section{Conclusion} \label{sec:conclusions}

In this paper we reviewed the co-design activities in the Center for Efficient
Exascale Discretizations of the Exascale Computing Project, focused on the
computational motif of PDE discretizations on general unstructured grids, with
emphasis on high-order methods. We described our co-design approach together
with the mathematical and algorithmic foundations for high-order finite element
discretizations. We also reviewed other topics that are necessary for a complete
high-order ecosystem, such as matrix-free linear solvers and high-order mesh
adaptivity, which are still active areas of research. A number of freely
available software products are being actively developed and supported by the
center, including benchmarks, miniapps and high- and low-level library APIs.
The CEED team is very much interested in collaborations and feedback, please
visit our website \url{ceed.exascaleproject.org} to get in touch.

%% file: tex/acknowledgments.tex
\section*{Acknowledgments}

This research is supported by the Exascale Computing Project (17-SC-20-SC), a
collaborative effort of two U.S. Department of Energy organizations (Office of
Science and the National Nuclear Security Administration) responsible for the
planning and preparation of a capable exascale ecosystem, including software,
applications, hardware, advanced system engineering and early testbed platforms,
in support of the nation's exascale computing imperative.

The research used resources of the Argonne Leadership Computing Facility, which
is supported by the U.S. Department of Energy, Office of Science, under Contract
DE-AC02-06CH11357. This research also used resources of the Oak Ridge Leadership
Computing Facility at Oak Ridge National Laboratory, which is supported by the
Office of Science of the U.S. Department of Energy under Contract DE-AC05-00OR22725.
Work performed under the auspices of the U.S. Department of Energy under
Contract DE-AC52-07NA27344 (LLNL-JRNL-814059).

%% file: tex/funding.tex
\section*{Funding}

This material is based upon work supported by the U.S. Department of Energy,
Office of Science, under Contracts DE-AC02-06CH11357, DE-AC05-00OR22725 and
DE-AC52-07NA27344.